\documentclass[english, reprint, superscriptaddress, aps, floatfix, prl]{revtex4-2}
\usepackage[T1]{fontenc}
\usepackage[utf8]{inputenc}

\newcommand{\mathdefault}[1][]{}

\usepackage{tabularx}

\usepackage[english]{babel}

\usepackage{amsmath,amsthm,amssymb,amsfonts}
\usepackage{mathtools}

\usepackage[dvipsnames, usenames]{xcolor}
\usepackage{xspace}

\colorlet{mylinkcolor}{RoyalPurple}
\colorlet{mycitecolor}{RoyalPurple}
\colorlet{myurlcolor}{RoyalPurple}
\usepackage{hyperref}
\hypersetup{
	linkcolor  = mylinkcolor,
	citecolor  = mycitecolor,
 	urlcolor   = myurlcolor,
 	colorlinks = true,
 	breaklinks = true
 }

\makeatletter

\pdfpageheight\paperheight
\pdfpagewidth\paperwidth

\makeatother

\usepackage{graphicx}
\usepackage[caption=false]{subfig}
\usepackage{import}

\usepackage{csquotes}
\MakeOuterQuote{"}

\usepackage[uncertainty-mode=separate, per-mode=symbol, group-digits=true]{siunitx}

\usepackage{bm}

\usepackage{xfrac}
\usepackage{letltxmacro}
\usepackage{braket}

\usepackage{pgf}

\newcommand{\SiN}{\ensuremath{\mathrm{Si_3 N_4}}\xspace}

\usepackage{hyphenat} 
\newcommand{\cauchy}{Cauchy\hyp{}Schwarz\xspace} 
\newcommand{\fperot}{Fabry\hyp{}Pérot\xspace}

\DeclareSIUnit{\dBc}{dBc}
\DeclareSIUnit{\ppm}{ppm}
\DeclareSIUnit{\wmk}{\W\per\m\per\K}

\usepackage{commath}

\usepackage{letltxmacro}
\LetLtxMacro{\oldsqrt}{\sqrt}
\renewcommand{\sqrt}[2][]{\oldsqrt[#1]{#2}\,}

\newcommand{\nbar}{\bar{n}}

\newcommand{\meff}{m_\text{eff}}
\newcommand{\Gopt}{\Gamma_\mathrm{opt}}
\newcommand{\GS}{\Gamma_\mathrm{S}}
\newcommand{\GAS}{\Gamma_\mathrm{AS}}
\newcommand{\Gm}{\Gamma_\mathrm{m}}
\newcommand{\Om}{\Omega_\mathrm{m}}
\newcommand{\nt}{\bar{n}_\mathrm{th}}
\newcommand{\ncav}{\bar{n}_\mathrm{cav}}

\newcommand{\Cq}{C_\mathrm{q}}

\newcommand{\gtwo}{g^{(2)}}

\newcommand{\gnot}{g_{0}}

\newcommand{\acr}{\hat{a}^{\dagger}}

\newcommand{\ban}{\hat{b}}

\newcommand{\bcr}{\hat{b}^{\dagger}}

\newcommand{\ham}[1]{\hat{H}_\text{#1}}

\newcommand{\harmconj}{\text{h.c.}}

\newcommand{\perc}[1]{\qty{#1}{\percent}}

\newcommand{\Eqref}[1]{{Eq.}~\ref{#1}}
\newcommand{\fref}[1]{{Fig.}~\ref{#1}}
\newcommand{\figuref}[1]{{Figure}~\ref{#1}}

\newcommand{\nv}{\varnothing}
\newcommand{\ov}{\circ}

\makeatother

\begin{document}
\title{Non-classical correlations between photons and phonons of center-of-mass motion of a mechanical oscillator} 

\newcommand{\nbi}{Niels Bohr Institute, University of Copenhagen, Blegdamsvej 17, 2100 Copenhagen, Denmark}

\author{Ivan \surname{Galinskiy}}
\affiliation{\nbi}

\author{Georg \surname{Enzian}}
\affiliation{\nbi}

\author{Michał \surname{Parniak}}
\affiliation{\nbi}
\affiliation{Faculty of Physics, University of Warsaw, L. Pasteura 5, 02-093 Warsaw, Poland}
\affiliation{Centre for Quantum Optical Technologies, Centre of New Technologies, University of Warsaw, S. Banacha 2c, 02-097 Warsaw, Poland}

\author{Eugene S. \surname{Polzik}}
\email{polzik@nbi.ku.dk}
\affiliation{\nbi}

\begin{abstract}
  We demonstrate non-classical correlations between phonons and photons created using opto-mechanical spontaneous parametric down-conversion in a system based on a soft-clamped ultracoherent membrane oscillator inside of a \fperot{} optical resonator. Non-Gaussian quantum features are demonstrated for the center-of-mass motion of a sub-millimeter nanogram-scale mechanical oscillator. We show that phonons stored in the mechanical oscillator, when subsequently read out, display strong signs of quantum coherence, which we demonstrate by single-photon counting enabled by our state-of-the-art optical filtering system. We observe a violation of the classical two-time \cauchy{} inequality between a heralding write photon and a stored phonon with a confidence of $> \perc{92}$. 
\end{abstract}

\maketitle

In recent years, there has been substantial progress towards increasingly macroscopic \cite{Nimmrichter2013} systems that display quantum properties. From collective spins of atomic gases and motion \cite{thomasEntanglementDistantMacroscopic2021}, both at room temperature and ultracold \cite{kargLightmediatedStrongCoupling2020}, through nano- and microscopic mechanical oscillators \cite{ockeloen-korppiStabilizedEntanglementMassive2018}, and recently with high-mass systems in bulk acoustic wave resonators \cite{bildSchrodingerCatStates2023, Schrinski2023}, quantum effects are now being observed in multi-\unit{\kg} systems in gravitational wave detectors \cite{acerneseQuantumBackactionKgScale2020New,whittleApproachingMotionalGround2021New}. Moreover, a combination of mechanical systems with quantum-level nonlinearities, such as superconducting qubits \cite{chuQuantumAcousticsSuperconducting2017,gustafssonPropagatingPhononsCoupled2014}, have allowed to exploit their non-classical behavior for quantum storage and processing. Recent results for microscopic oscillators with frequencies in \unit{\THz} \cite{leeMacroscopicNonclassicalStates2012} and \unit{\GHz} domains have pushed the boundary of nongaussianity in optomechanical systems, with DLCZ-based \cite{Duan2001} single-phonon protocols \cite{hongHanburyBrownTwiss2017, Riedinger2016}, and state tomography \cite{bildSchrodingerCatStates2023, enzianNonGaussianMechanicalMotion2021}. 

A special challenge is to observe nonclassical states of motion of center-of-mass macroscopic system which is of fundamental interest \cite{Belenchia2018}. Quantum fluctuations of a ground state can be seen in massive systems \cite{acerneseQuantumBackactionKgScale2020New}, but it is more challenging to prepare macroscopic mechanical systems in demonstrably nonclassical and nongaussian states. 
In this Letter, we report the observation of nonclassical single photon - single phonon correlated states where phonons belong to the center-of-mass motion of a mm-size mechanical oscillator. 
The oscillator used in our work is visible to the naked eye and, maybe more importantly, shows the potential to break into the uncharted territory of macroscopicity for quantum-mechanical behavior \cite{Nimmrichter2013}.

The mechanical oscillator is a hexagonal defect with a diameter of \qty{0.3}{\mm} in a phononic crystal embedded in a membrane, with frequency of ${\Om \approx 2\pi \times \qty{1.4}{\MHz}}$, and effective mass of $\meff \approx \qty{2}{\nano\gram}$, which places us in the low-frequency, high-mass regime. Observing such behavior requires precise control over most of our experimental parameters, which we demonstrate in the domains of noise, stability, and interaction tuning. By showing a significant violation of the \cauchy{} inequality in the photon statistics, we demostrate nongaussian behavior in a low-frequency, macroscopic purely optomechanical system.

\begin{figure}[pb]
  \centering
  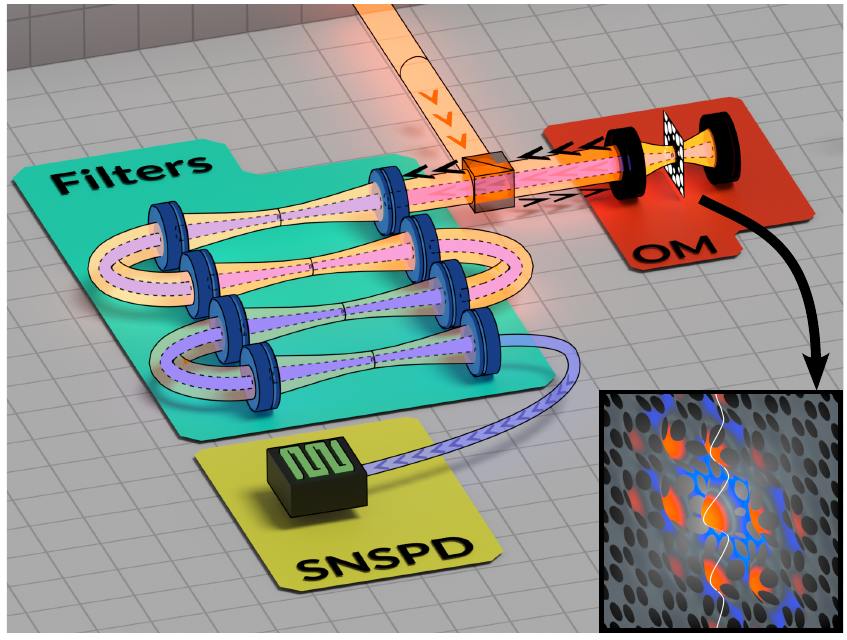
  \caption{\label{fig:main-setup-diagram}Overview of the experimental setup. The optomechanical system ("OM") consists of a mechanical membrane oscillator, with the fundamental mode shown in the bottom-right inset,  placed inside a high-Q \fperot optical cavity. The input pump light (coming from the center top) interacts with the mechanical system. The scattered Stokes (write) or anti-Stokes (read) photons are separated from the pump light with a set of filter cavities ("Filters") and are detected by a superconducting nanowire single photon detector ("SNSPD"). }
  
\end{figure}

The square \SiN{} membrane in which the oscillator is embedded has an overall size of \qtyproduct{3 x 3}{\mm} and a thickness of \qty{13}{\nm}. The majority of its area is occupied by a phononic shield that surrounds the defect where the high-Q mode of oscillation is mostly localized \cite{Tsaturyan2017} (\fref{fig:main-setup-diagram}). The phononic shield has a vibrational bandgap centered on the frequency of the main mode. The shield prevents coupling between the modes of the defect and the surroundings, resulting in Q factors above \num{200e6} at cryostat operating temperatures of \qtyrange{6}{9}{\K}.

The phononic-crystal membrane is positioned in free space between two mirrors of a \fperot resonator (membrane in the middle - MIM) operating at a wavelength of \qty{852}{\nm}. Mechanical motion of the dielectric membrane shifts the resonance frequency of this resonator which is a fundamentally nonlinear process of optomechanical coupling. 
We prevent environmental vibrations from affecting the stability of our system by suspending the cavity assembly on four steel springs, with a copper braid providing a thermal link between the OM cavity and the cryostat (see SI). Two superpolished concave mirrors with a ROC of \qty{25}{\mm} \footnote{Fabricated by FiveNine Optics}, form a cavity with an intrinsic finesse of approximately \num{11e3}. With a resonator length of \qty{18}{\mm}, the resulting optical bandwidth $\kappa$ is \qtyrange{0.7}{1}{\MHz}, with the precise value defined by the relative position between the standing optical wave and the membrane.

Nonclassical motional states are generated by conditional (heralded) mechanical state preparation, where the heralding comes from single-photon detection, as theoretically discussed in \cite{PhysRevLett.107.123601,PhysRevLett.110.010504,{galland}}
and experimentally demonstrated in atomic \cite{dideriksenRoomtemperatureSinglephotonSource2021a, corzoWaveguidecoupledSingleCollective2019}, and high-frequency optomechanical systems \cite{hongHanburyBrownTwiss2017,leeMacroscopicNonclassicalStates2012}. The sideband-resolved operation allows us to change the kind of optomechanical interaction by either red- or blue-detuning the pumping laser field with respect to the optical cavity resonance.

\begin{figure}[tb]
  \centering
  \renewcommand\sffamily{}
  \newcommand\hzz{\Hz}
    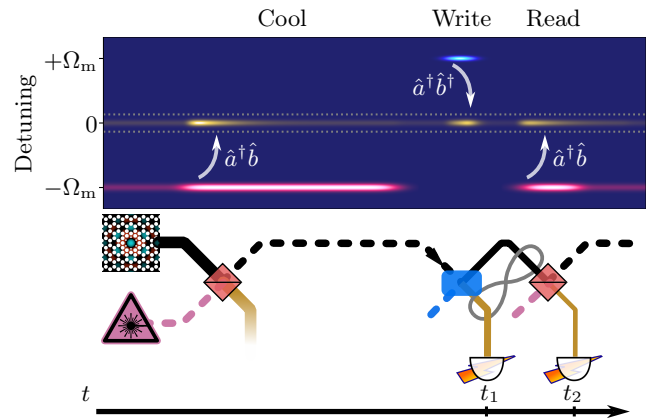
  \caption{\label{fig:full-protocol-schematic}Schematic representation of the protocol used in the experiment. The top plot is a spectrogram illustrating the timing and frequencies of optical drives entering the optomechanical cavity (red and blue traces), and the generation of scattered photons resonant with the cavity (yellow). The bottom drawing illustrates the equivalent "optical diagram" of the sequence, consisting of a beamsplitter - SPDC - beamsplitter chain of interactions spread over time. Note that the two detection events at times $t_{1,2}$ are sufficiently separated in time, such that only one single-photon detector is needed for their detection. See the main text for a detailed description of the interactions.}
  
\end{figure}

Our protocol utilizes basic steps of a quantum repeater protocol \cite{Duan2001}, where we begin by creating low gain two-mode squeezing between the mechanical and optical modes using the "write" pulse, as shown schematically in \fref{fig:full-protocol-schematic}. As a result, a photon in the propagating optical mode detected by a single-photon counter heralds the creation of a phonon of mechanical excitation. The mechanical mode, localized to the membrane, is stored for a chosen time delay, after which it can be converted back into the optical domain by applying a read pulse. Effectively, this leads to entangled photon-phonon modes, which result in two entangled time-separated optical modes. Correlations between these two modes are the primary indicator of nonclassical behavior in our system.

The basis of our preparation and measurement scheme is the sideband-resolved operation of our optomechanical system, as its optical decay rate (cavity linewidth) $\kappa \approx 2\pi \times \qty{0.8}{\MHz}$ is significantly smaller than the mechanical frequency $\Om \approx 2\pi \times \qty{1.4}{\MHz}$. In the scheme described in \cite{galland}, this leads to two kinds of interplay between mechanical motion and optical cavity mode: beamsplitter interaction and two-mode squeezing interaction, to which we will refer as spontaneous parametric downconversion or SPDC. Beamsplitter interaction, or anti-Stokes scattering, performs state swapping between light and mechanics. This mode of operation is achieved when the pump field is detuned to the red side of the optomechanical cavity's optical resonance ($\Delta = -\Om$) \cite{Aspelmeyer2014}, favoring swapping of the motional state  onto the optical state. SPDC interaction, on the other hand, leads to Stokes scattering, or the creation of phonon-photon pairs achieved by detuning the pump field to the blue side of the resonance ($\Delta = +\Om$).

As the main tool of our investigations of mechanical state properties, we use photon counting statistics, specifically second-order correlations \cite{reidViolationsClassicalInequalities1986} between photon counts defined as  $\gtwo_{12} \equiv \braket{n_1 n_2}/\braket{n_1} \braket{n_2}$, where $n_{1,2}$ is the photon number for modes 1 or 2, respectively. The modes can be e.g. physically different light beams (geometrical modes), or time-separated detection windows (temporal modes). Our experiment uses photon counting in temporal modes, i.e. numbers of photons detected in time windows around $t_1$ and $t_2$.

Correlations between scattered photons from the write and read pulses 
reflect correlations between the optical and mechanical state after a period of free evolution. The photon statistics can be used as a proxy to determine the classicality of a system \cite{reidViolationsClassicalInequalities1986}. More precisely, in our case a violation of the \cauchy{} inequality demonstrates nonclassical correlations between two optical modes at times $t_1$ and $t_2$. The inequality, when applied to classical states of light (for which the Glauber–Sudarshan P function has the properties of a probability distribution), yields
\begin{equation}\label{eq:cauchy_raw}
    \norm{\gtwo_{12}}^2 \leq \gtwo_1 \cdot \gtwo_2,
\end{equation}
where we defined a shorthand notation for the two-time (two-mode) correlation $\gtwo_{12} \equiv \gtwo(t_1, t_2)$, and autocorrelation $\gtwo_{n} \equiv \gtwo(t_n, t_n)$. It is convenient to define the \emph{\cauchy parameter}, that we will denote as $R$, such that
\begin{equation}\label{eq:R}
    R \equiv \frac{\norm{\gtwo_{12}}^2}{\gtwo_1 \cdot \gtwo_2}.
\end{equation}
Therefore, if $R > 1$, then the \cauchy inequality of \Eqref{eq:cauchy_raw} is not applicable, indicating non-classical photon statistics, where $P(\alpha)$ can no longer be interpreted as a probability distribution.
Note that the knowledge of $\gtwo_{12}$ per se is not enough to guarantee non-classicality, as there exist classical sources with strong bunching ($\gtwo_1 \gg 1$). It is only the \cauchy parameter that can establish non-classicality in this case.

\begin{figure}[tpb]
  \centering
  \renewcommand\sffamily{}
  \newcommand\hzz{\Hz}
        \includegraphics[width=1\columnwidth]{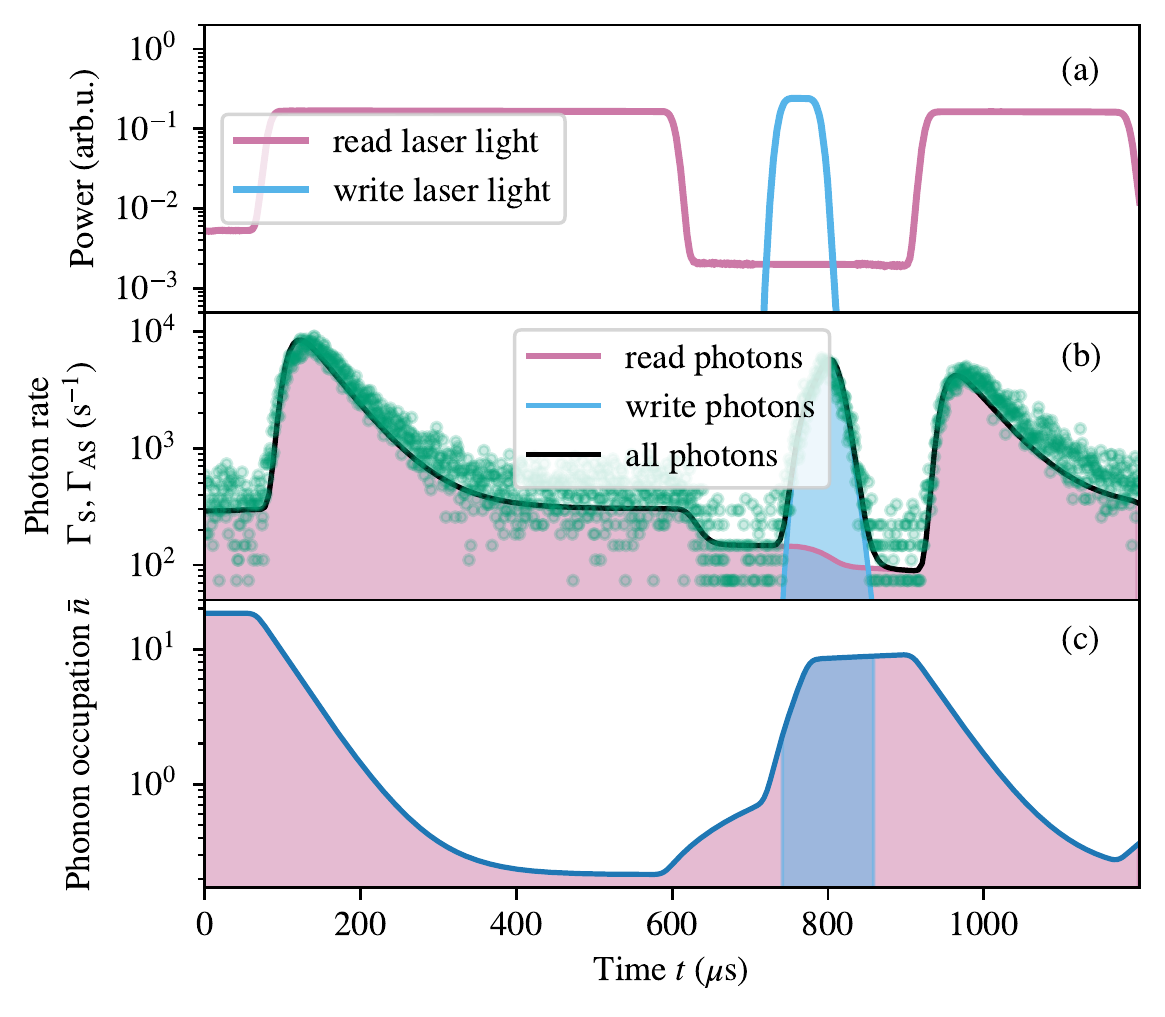}
  \caption{\label{fig:exp-high-rate} Photon rates averaged over many experimental "shots", showing driving laser fields (a), photon rates (b), and phonon occupation estimated from a model (c). Points represent the raw experimental result, while solid lines represent the model outcome, as described in the Supplementary Information \cite{suppmat}. The red color corresponds to cooling and reading out the excitation, and the blue color shows the writing process. The solid black line in (b) is the total photon rate, which is a sum of red (read/anti-Stokes photons, $\GAS$) and blue lines (write/Stokes photons, $\GS$).} The excitation probability has been increased for better visual clarity. Note that the photon rates are delayed with respect to the respective laser pulses due to the time-domain response of filtering cavities.
  
\end{figure}

To reiterate, our experiment measures correlations between photons detected during two time-separated pulses, which we call "write" and "read" respectively. During writing, the optomechanical system undergoes a process equivalent to SPDC. There, pump photons are scattered down in energy, while creating excitations in the mechanical oscillator, leading to entangled phonon-photon pairs, corresponding to the interaction Hamiltonian $\ham{int} \propto \acr \bcr + \harmconj$. In the low-gain regime relevant for our case, this interaction leads to a state $\ket{0,0} + \alpha \ket{1,1}$ with the first (second) index corresponding to the photon (phonon) number, and the probability amplitude being $\alpha \ll 1$. 

While the created phonon remains localized in the mechanical oscillator, the photon can be detected externally. In what is rather similar to the DLCZ protocol, the detected photon serves as a herald of the scattering event. As photodetection is a fundamentally nonlinear process, detection of this photon under appropriate conditions (discussed further down) is capable of collapsing the entangled mechanics-light state onto a mechanical single-phonon Fock state.

During "reading", the interaction changes to that of a beamsplitter, where the intracavity optical state is swapped with the mechanical state, to a degree controlled by the interaction strength (broadening) $\Gopt$ and duration ($T_r$) of the reading pulse. Particularly in the sideband-resolved regime, $\Gopt = 4 \gnot^2 \ncav/\kappa$, where $\gnot$ is the single-photon optomechanical coupling rate, $\kappa$ is the optical cavity linewidth, and $\ncav$ is the intracavity photon number of the pumping field.

A prominent feature of our experiment is that we operate with a macroscopic oscillator and, as opposed, for example, to other experiments with macroscopic objects in non-Gaussian nonclassical states \cite{bildSchrodingerCatStates2023, Schrinski2023}, we study the center-of-mass mechanical motion. This results in a relatively low mechanical frequency of the oscillator which, in turn, constitutes a significant challenge for the detection of single photons at the mechanical sideband frequencies. The frequency difference between the sidebands created by the mechanical motion and the base optical frequency is only \qty{1.4}{\MHz}, corresponding to a fractional frequency difference of less than \num{4e-9}. To achieve suppression of the unscattered pump light and neighboring mechanical modes, we designed and built a state-of-the-art filter system. The filtering is performed by an array of four \qty{60}{\cm} \fperot resonators \cite{ourOptica}, positioned in series along the path of light that has interacted with the optomechanical system (\fref{fig:main-setup-diagram}). The individual linewidth of each filter resonator is approximately \qty{30}{\kHz}, chosen to be broad enough to allow the optically broadened mechanical signal to be transmitted, while efficiently blocking the unscattered pump light, and scattering by nearby out-of-bandgap mechanical modes.

Classical noise in low-frequency systems, such as ours, is a big concern, and is challenging to counteract. Some of the largest sources of classical fluctuations are phase noise coming from our laser system, thermal motion of mirrors in the optomechanical cavity, and electronic noise transduced into the light via e.g. electrooptic modulators. All these sources contribute to an increase of phase noise in the light entering the system. Especially in the context of sideband-resolved operation, phase noise is transduced into intensity noise inside the optomechanical cavity, leading to force noise on the mechanics. Additionally, optical noise resonant with the optomechanical cavity contributes spurious photon counts, which worsens the signal-to-noise ratio during single-photon counting.

We suppress the phase noise of our laser source using an active feedback based on a fiber interferometer \cite{delaylinepaper}. Electronic noise from RF drives is filtered using custom high-Q LC filters. Mirror noise is minimized by choosing a suitable geometry of cavity mirrors \cite{jonasMScThesis,suppmat}. Finally, as cooling down deeply into the ground state is essential, for cooling we use the laser directly (as the carrier), without any modulation. This way we avoid introducing electronic noise during cooling and read-out. The electro-optic modulator is then used to either extinguish the carrier drive and/or generate the blue-detuned drive. 

The experiment runs in a pulsed regime, with each of the experimental "shots" (roughly \qty{1}{\ms} each) consisting of three main stages: ground-state cooling of the mechanical oscillator, SPDC-like creation of a photon-phonon pair, and readout of the created phonon onto a photon. We shall refer to these steps, schematically shown in \fref{fig:full-protocol-schematic}, as "cool", "write", and "read", respectively.

\begin{figure*}[tb]
    \centering
    \renewcommand\sffamily{}
    
    \includegraphics[width=0.95\textwidth]{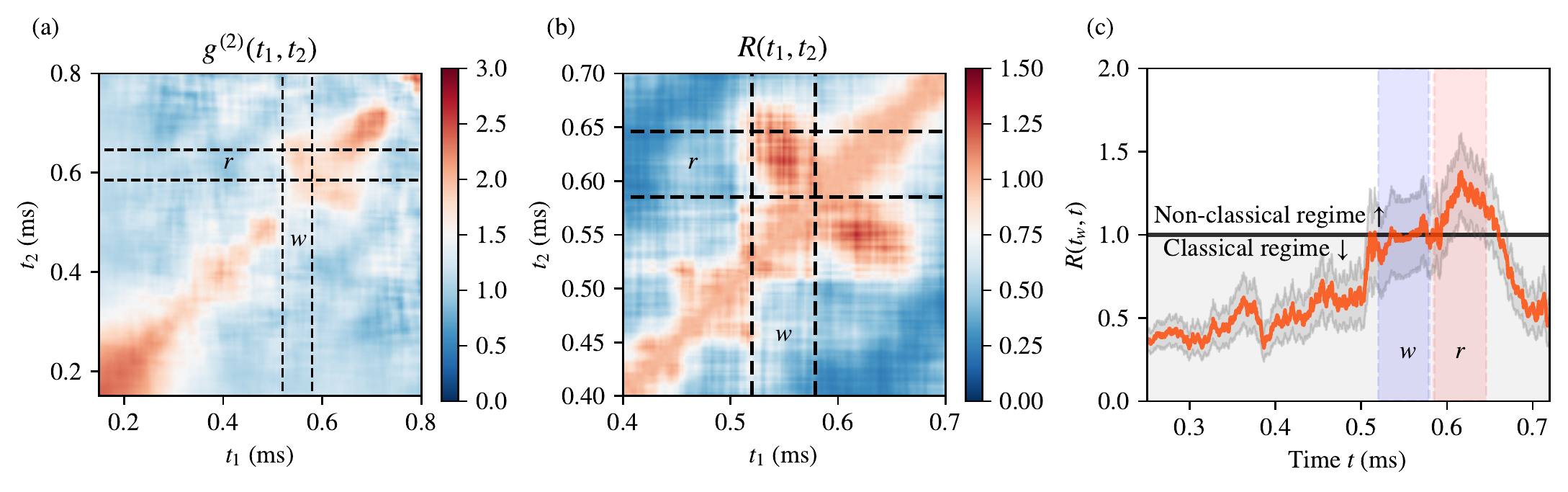}
    \caption{\label{fig:exp-cauchy}Experimental data presenting non-classical values of the correlation function $\gtwo$ in panel (a) (\Eqref{eq:cauchy_raw}) and the \cauchy parameter $R$ in panel (b) (\Eqref{eq:R}).The off-diagonal lobes in (a) and (b) outlined by the dashed lines show the non-classical regions of the distributions. Panel (c) presents an optimal section of map (b), where the write window position $t_w$ is kept constant in the middle of region $w$ denoted by dashed lines in panel (b), and the read position is varied. The red curve is the mean value while shaded gray regions represent one standard deviation.}
  
\end{figure*}

The cooling pulse is necessary to bring the membrane's main mode motion as close to the ground state as possible, as the residual thermal occupation has detrimental effects on the rest of the experiment \cite{galland}. In thermal equilibrium and in the absence of any optical interaction, the occupation stabilizes at an approximate level of $\nt = k T_\text{env} / (\hbar \Om) = \num{200(20)e3}$, with $T_\text{env}\approx\qty{15}{\K}$ corresponding to the assembly temperature of about \qtyrange{6}{9}{\K}, as witnessed in previous works with almost identical systems \cite{thomasEntanglementDistantMacroscopic2021,galinskiyPhononCountingThermometry2020}. The 10\% uncertainty represents typical errors found in past works for this calibration. The corresponding thermal decoherence rate is $\Gamma_\text{th} \equiv \Gm \nt = 2\pi \cdot \qty{1.5(0.2)}{\kHz}$. Activation of the cooling pulse quickly brings the occupation down to \num{0.25 \pm 0.05} phonons. Notably, this estimate is based on a simulation and is not a direct measurement, however, its main purpose is to give an understanding of system dynamics, as the final correlation results are independent of this estimate. This level is set by the finite quantum cooperativity $\Cq \approx 10$ of our system, and various sources of technical noise, e.g. mirror thermal noise, laser phase noise, etc.

We note that during idle time between experimental shots, we keep the cooling beam active, though at a small fraction of the "read" power ($\approx \perc{1}$). This precooling maintains the phononic occupation at a relatively low level of about \num{20}, shortening the ground-state cooling time, while maintaining a low average optical power. The latter prevents spurious heating of the optomechanical cavity mirrors, for which we choose an idle time significantly longer than each experimental shot, resulting in a shot/idle duty cycle of \perc{5}.

After the occupation is reduced by the cooling, the red-detuned beam is shut off and a blue-detuned beam is turned on to enable the SPDC interaction creating phonon-photon pairs, parametrically amplifying the optomechanical state. Due to technical imperfections an insignificant red-detuned drive power remains present in the system. While this interaction is also equivalent to two-mode squeezing, single-photon detection introduces the nonlinearity necessary for nongaussian state preparation. Similar to the cooling pulse, we define the interaction strength as $g_w = 2 \gnot^2 \ncav/\kappa$. This interaction creates an average of $2g_w T_w$ phonon-photon pairs in the limit of low excitation \cite{galland} where $T_w$ is the pulse duration.  $T_w$ and $g_w$ are set such that the probability of scattering is low enough, in accordance with our heralded probabilistic scheme.

After the writing pulse, the mechanical state is read out onto light. Due to finite quantum cooperativity $\Cq \equiv \Gopt / \Gamma_\text{th} \approx 10$, the readout signal will necessarily contain a small admixture of thermal phonons from the environment given by $1/\Cq \approx \num{0.1}$. In practice, we make the readout pulse sufficiently long to achieve a steady state to guarantee that a full state swap has been achieved ($\Gopt T_r \gg 1$). Afterwards, a suitable detection window is chosen during postprocessing. Losses due to non-unity cavity extraction efficiency (overcoupling) ($\approx \qty{2}{\dB}$), losses in the filter cavity system ($\approx \qty{5}{\dB}$), fiber interconnects ($\approx \qty{3}{\dB}$), and downstream light propagation and detection ($\approx \qty{1}{\dB}$), limit our overall collection efficiency to approximately \qtyrange{5}{10}{\percent}. Fortunately, apart from a longer necessary acquisition time, two-photon correlations are not degraded by optical losses. In the end, the read-out phononic mode converted into a light mode propagates to the single-photon counter, where it is detected, serving its role as the second mode in two-mode correlations.

As shown in the bottom half of \fref{fig:full-protocol-schematic}, the whole experimental sequence can be graphically represented as an optical diagram, noting that the black-colored "phonon beam propagation" is, in fact, propagation in time. The interaction between the light and mechanics, shown in the top half of \fref{fig:full-protocol-schematic}, is represented as an "optomechanical spectrogram", where the optical drives (red and blue lines) are scattered by their corresponding interaction Hamiltonians. The scattered photons, shown in yellow, are resonant with the optomechanical cavity and are transmitted by our filtering system.

A typical experimental run consists of many "shots", usually in the range of millions, to overcome our moderate total photon collection efficiency (\perc{5}). \figuref{fig:exp-high-rate} shows an example of resulting photon rates, where the generation rate is increased for visual clarity. One can see the cool-write-read sequence averaged over 150 thousand experimental shots. The driving field strength is shown in the top graph. Cooling is happening between \qtyrange{150}{650}{\us}, followed by the "write" pulse between \qtyrange{750}{850}{\us} , and the "read" pulse starting around \qty{950}{\us}. The resulting photon count rates ($\GS$ and $\GAS$ for Stokes and anti-Stokes, respectively) are shown in the middle plot. With the knowledge of dynamics of photon rates in our system, we extract the mechanical occupation using a fit to photon rates, as shown in the bottom graph of \fref{fig:exp-high-rate} and described extensively in the Supplementary Information \cite{suppmat}.

In the low-probability excitation regime ($g_w T_w \approx 0.1$), we obtain correlation maps shown in \fref{fig:exp-cauchy}. After accumulating the equivalent of about \num{200} coincidences, we performed data analysis of the events. This included taking proper time subsections of the whole data-trace to maximize the overlap with temporal modes. We select temporal modes as shown in the figure by dashed lines, and calculate both the $\gtwo(t_1,t_2)$ and $R(t_1,t_2)$ under this constraint by shifting the mode positions to $t_1$, $t_2$. 

The $\gtwo$ map clearly shows nondiagonal elements with the value at optimal point $\gtwo_{wr}=1.89\pm0.13$ suggesting strong correlations between $r$ and $w$ photons. We also observe autocorrelation of thermal photons scattered in both processes as the diagonal feature in the maps, and $\gtwo_{ww}=1.54\pm0.15$ and $\gtwo_{rr}=1.67\pm0.09$ at optimal time windows. The nonclassicality is revealed in the enlarged map of $R$ in \fref{fig:exp-cauchy}, where at optimal mode positions we obtained the \cauchy parameter of $R = \num{1.37(26)}$, which violates the classical limit of $R = 1$. The corresponding probability of violation of the \cauchy inequality is $>\perc{92}$ (p-value of \num{0.08}). See also supplementary information for the correlation maps in the high-probability write case, which demonstrates strong albeit classical correlations.

Possible avenues for further improvements involve an increase in overall efficiency. Our losses are not dominated by any single source but are rather distributed along the many components in the light path. Notably, the system of filter cavities in its current state has a signal transmission of approximately \perc{30}, with the rest of the losses being mostly spread across fiber interconnects, mode matching, and internal losses of the optomechanical cavity. The strength of the optomechanical interaction, proportional to $\gnot$, can be increased by a more resilient system of mounting the mirrors and the membrane with respect to each other, which would increase the optomechanical mode overlap. Nevertheless, even with the limitations of our current setup, our results suggest that mechanical nonclassicality is achievable with photon counting even in such a low-frequency and large-size mechanical system with modest cooling. Further improvements, including heralded preparation of single-phonon states, are within reach for optomechanics with macroscopic membrane oscillators.

\begin{acknowledgments} 
This research has been funded by the European Research Council (ERC) Advanced Grant QUANTUM-N under the EU Horizon 2020 (grant agreement no. 787520), by Villum Fonden under a Villum Investigator Grant, grant no. 25880, and by the Novo Nordisk Foundation Center for Biomedical Quantum Sensing, grant NNF24SA0088433. G.E. acknowledges support from the European Union’s Horizon 2020 research and innovation programme under the Marie Sklodowska-Curie grant agreement no. 847523.
\end{acknowledgments}

\bibliography{refs,supplemental}

\begin{thebibliography}{31}%
\makeatletter
\providecommand \@ifxundefined [1]{%
 \@ifx{#1\undefined}
}%
\providecommand \@ifnum [1]{%
 \ifnum #1\expandafter \@firstoftwo
 \else \expandafter \@secondoftwo
 \fi
}%
\providecommand \@ifx [1]{%
 \ifx #1\expandafter \@firstoftwo
 \else \expandafter \@secondoftwo
 \fi
}%
\providecommand \natexlab [1]{#1}%
\providecommand \enquote  [1]{``#1''}%
\providecommand \bibnamefont  [1]{#1}%
\providecommand \bibfnamefont [1]{#1}%
\providecommand \citenamefont [1]{#1}%
\providecommand \href@noop [0]{\@secondoftwo}%
\providecommand \href [0]{\begingroup \@sanitize@url \@href}%
\providecommand \@href[1]{\@@startlink{#1}\@@href}%
\providecommand \@@href[1]{\endgroup#1\@@endlink}%
\providecommand \@sanitize@url [0]{\catcode `\\12\catcode `\$12\catcode
  `\&12\catcode `\#12\catcode `\^12\catcode `\_12\catcode `\%12\relax}%
\providecommand \@@startlink[1]{}%
\providecommand \@@endlink[0]{}%
\providecommand \url  [0]{\begingroup\@sanitize@url \@url }%
\providecommand \@url [1]{\endgroup\@href {#1}{\urlprefix }}%
\providecommand \urlprefix  [0]{URL }%
\providecommand \Eprint [0]{\href }%
\providecommand \doibase [0]{https://doi.org/}%
\providecommand \selectlanguage [0]{\@gobble}%
\providecommand \bibinfo  [0]{\@secondoftwo}%
\providecommand \bibfield  [0]{\@secondoftwo}%
\providecommand \translation [1]{[#1]}%
\providecommand \BibitemOpen [0]{}%
\providecommand \bibitemStop [0]{}%
\providecommand \bibitemNoStop [0]{.\EOS\space}%
\providecommand \EOS [0]{\spacefactor3000\relax}%
\providecommand \BibitemShut  [1]{\csname bibitem#1\endcsname}%
\let\auto@bib@innerbib\@empty
\bibitem [{\citenamefont {Nimmrichter}\ and\ \citenamefont
  {Hornberger}(2013)}]{Nimmrichter2013}%
  \BibitemOpen
  \bibfield  {author} {\bibinfo {author} {\bibfnamefont {S.}~\bibnamefont
  {Nimmrichter}}\ and\ \bibinfo {author} {\bibfnamefont {K.}~\bibnamefont
  {Hornberger}},\ }\href {https://doi.org/10.1103/PhysRevLett.110.160403}
  {\bibfield  {journal} {\bibinfo  {journal} {Phys. Rev. Lett.}\ }\textbf
  {\bibinfo {volume} {110}},\ \bibinfo {pages} {160403} (\bibinfo {year}
  {2013})}\BibitemShut {NoStop}%
\bibitem [{\citenamefont {Thomas}\ \emph {et~al.}(2021)\citenamefont {Thomas},
  \citenamefont {Parniak}, \citenamefont {{\O}stfeldt}, \citenamefont
  {M{\o}ller}, \citenamefont {B{\ae}rentsen}, \citenamefont {Tsaturyan},
  \citenamefont {Schliesser}, \citenamefont {Appel}, \citenamefont {Zeuthen},\
  and\ \citenamefont {Polzik}}]{thomasEntanglementDistantMacroscopic2021}%
  \BibitemOpen
  \bibfield  {author} {\bibinfo {author} {\bibfnamefont {R.~A.}\ \bibnamefont
  {Thomas}}, \bibinfo {author} {\bibfnamefont {M.}~\bibnamefont {Parniak}},
  \bibinfo {author} {\bibfnamefont {C.}~\bibnamefont {{\O}stfeldt}}, \bibinfo
  {author} {\bibfnamefont {C.~B.}\ \bibnamefont {M{\o}ller}}, \bibinfo {author}
  {\bibfnamefont {C.}~\bibnamefont {B{\ae}rentsen}}, \bibinfo {author}
  {\bibfnamefont {Y.}~\bibnamefont {Tsaturyan}}, \bibinfo {author}
  {\bibfnamefont {A.}~\bibnamefont {Schliesser}}, \bibinfo {author}
  {\bibfnamefont {J.}~\bibnamefont {Appel}}, \bibinfo {author} {\bibfnamefont
  {E.}~\bibnamefont {Zeuthen}},\ and\ \bibinfo {author} {\bibfnamefont {E.~S.}\
  \bibnamefont {Polzik}},\ }\href {https://doi.org/10.1038/s41567-020-1031-5}
  {\bibfield  {journal} {\bibinfo  {journal} {Nature Physics}\ }\textbf
  {\bibinfo {volume} {17}},\ \bibinfo {pages} {228} (\bibinfo {year}
  {2021})}\BibitemShut {NoStop}%
\bibitem [{\citenamefont {Karg}\ \emph {et~al.}(2020)\citenamefont {Karg},
  \citenamefont {Gouraud}, \citenamefont {Ngai}, \citenamefont {Schmid},
  \citenamefont {Hammerer},\ and\ \citenamefont
  {Treutlein}}]{kargLightmediatedStrongCoupling2020}%
  \BibitemOpen
  \bibfield  {author} {\bibinfo {author} {\bibfnamefont {T.~M.}\ \bibnamefont
  {Karg}}, \bibinfo {author} {\bibfnamefont {B.}~\bibnamefont {Gouraud}},
  \bibinfo {author} {\bibfnamefont {C.~T.}\ \bibnamefont {Ngai}}, \bibinfo
  {author} {\bibfnamefont {G.-L.}\ \bibnamefont {Schmid}}, \bibinfo {author}
  {\bibfnamefont {K.}~\bibnamefont {Hammerer}},\ and\ \bibinfo {author}
  {\bibfnamefont {P.}~\bibnamefont {Treutlein}},\ }\href
  {https://doi.org/10.1126/science.abb0328} {\bibfield  {journal} {\bibinfo
  {journal} {Science}\ }\textbf {\bibinfo {volume} {369}},\ \bibinfo {pages}
  {174} (\bibinfo {year} {2020})}\BibitemShut {NoStop}%
\bibitem [{\citenamefont {{Ockeloen-Korppi}}\ \emph {et~al.}(2018)\citenamefont
  {{Ockeloen-Korppi}}, \citenamefont {Damsk{\"a}gg}, \citenamefont
  {Pirkkalainen}, \citenamefont {Asjad}, \citenamefont {Clerk}, \citenamefont
  {Massel}, \citenamefont {Woolley},\ and\ \citenamefont
  {Sillanp{\"a}{\"a}}}]{ockeloen-korppiStabilizedEntanglementMassive2018}%
  \BibitemOpen
  \bibfield  {author} {\bibinfo {author} {\bibfnamefont {C.~F.}\ \bibnamefont
  {{Ockeloen-Korppi}}}, \bibinfo {author} {\bibfnamefont {E.}~\bibnamefont
  {Damsk{\"a}gg}}, \bibinfo {author} {\bibfnamefont {J.-M.}\ \bibnamefont
  {Pirkkalainen}}, \bibinfo {author} {\bibfnamefont {M.}~\bibnamefont {Asjad}},
  \bibinfo {author} {\bibfnamefont {A.~A.}\ \bibnamefont {Clerk}}, \bibinfo
  {author} {\bibfnamefont {F.}~\bibnamefont {Massel}}, \bibinfo {author}
  {\bibfnamefont {M.~J.}\ \bibnamefont {Woolley}},\ and\ \bibinfo {author}
  {\bibfnamefont {M.~A.}\ \bibnamefont {Sillanp{\"a}{\"a}}},\ }\href
  {https://doi.org/10.1038/s41586-018-0038-x} {\bibfield  {journal} {\bibinfo
  {journal} {Nature}\ }\textbf {\bibinfo {volume} {556}},\ \bibinfo {pages}
  {478} (\bibinfo {year} {2018})}\BibitemShut {NoStop}%
\bibitem [{\citenamefont {Bild}\ \emph {et~al.}(2023)\citenamefont {Bild},
  \citenamefont {Fadel}, \citenamefont {Yang}, \citenamefont {{von L{\"u}pke}},
  \citenamefont {Martin}, \citenamefont {Bruno},\ and\ \citenamefont
  {Chu}}]{bildSchrodingerCatStates2023}%
  \BibitemOpen
  \bibfield  {author} {\bibinfo {author} {\bibfnamefont {M.}~\bibnamefont
  {Bild}}, \bibinfo {author} {\bibfnamefont {M.}~\bibnamefont {Fadel}},
  \bibinfo {author} {\bibfnamefont {Y.}~\bibnamefont {Yang}}, \bibinfo {author}
  {\bibfnamefont {U.}~\bibnamefont {{von L{\"u}pke}}}, \bibinfo {author}
  {\bibfnamefont {P.}~\bibnamefont {Martin}}, \bibinfo {author} {\bibfnamefont
  {A.}~\bibnamefont {Bruno}},\ and\ \bibinfo {author} {\bibfnamefont
  {Y.}~\bibnamefont {Chu}},\ }\href {https://doi.org/10.1126/science.adf7553}
  {\bibfield  {journal} {\bibinfo  {journal} {Science}\ }\textbf {\bibinfo
  {volume} {380}},\ \bibinfo {pages} {274} (\bibinfo {year}
  {2023})}\BibitemShut {NoStop}%
\bibitem [{\citenamefont {Schrinski}\ \emph {et~al.}(2023)\citenamefont
  {Schrinski}, \citenamefont {Yang}, \citenamefont {von L\"upke}, \citenamefont
  {Bild}, \citenamefont {Chu}, \citenamefont {Hornberger}, \citenamefont
  {Nimmrichter},\ and\ \citenamefont {Fadel}}]{Schrinski2023}%
  \BibitemOpen
  \bibfield  {author} {\bibinfo {author} {\bibfnamefont {B.}~\bibnamefont
  {Schrinski}}, \bibinfo {author} {\bibfnamefont {Y.}~\bibnamefont {Yang}},
  \bibinfo {author} {\bibfnamefont {U.}~\bibnamefont {von L\"upke}}, \bibinfo
  {author} {\bibfnamefont {M.}~\bibnamefont {Bild}}, \bibinfo {author}
  {\bibfnamefont {Y.}~\bibnamefont {Chu}}, \bibinfo {author} {\bibfnamefont
  {K.}~\bibnamefont {Hornberger}}, \bibinfo {author} {\bibfnamefont
  {S.}~\bibnamefont {Nimmrichter}},\ and\ \bibinfo {author} {\bibfnamefont
  {M.}~\bibnamefont {Fadel}},\ }\href
  {https://doi.org/10.1103/PhysRevLett.130.133604} {\bibfield  {journal}
  {\bibinfo  {journal} {Phys. Rev. Lett.}\ }\textbf {\bibinfo {volume} {130}},\
  \bibinfo {pages} {133604} (\bibinfo {year} {2023})}\BibitemShut {NoStop}%
\bibitem [{\citenamefont {Acernese}\ \emph {et~al.}(2020)\citenamefont
  {Acernese} \emph {et~al.}}]{acerneseQuantumBackactionKgScale2020New}%
  \BibitemOpen
  \bibfield  {author} {\bibinfo {author} {\bibnamefont {Acernese}} \emph
  {et~al.} (\bibinfo {collaboration} {The Virgo Collaboration}),\ }\href
  {https://doi.org/10.1103/PhysRevLett.125.131101} {\bibfield  {journal}
  {\bibinfo  {journal} {Physical Review Letters}\ }\textbf {\bibinfo {volume}
  {125}},\ \bibinfo {pages} {131101} (\bibinfo {year} {2020})}\BibitemShut
  {NoStop}%
\bibitem [{\citenamefont {Whittle}\ \emph {et~al.}(2021)\citenamefont
  {Whittle}, \citenamefont {Hall}, \citenamefont {Dwyer}, \citenamefont
  {Mavalvala}, \citenamefont {Sudhir}, \citenamefont {Abbott}, \citenamefont
  {Ananyeva}, \citenamefont {Austin}, \citenamefont {Barsotti}, \citenamefont
  {Betzwieser}, \citenamefont {Blair}, \citenamefont {Brooks}, \citenamefont
  {Brown}, \citenamefont {Buikema}, \citenamefont {Cahillane} \emph
  {et~al.}}]{whittleApproachingMotionalGround2021New}%
  \BibitemOpen
  \bibfield  {author} {\bibinfo {author} {\bibfnamefont {C.}~\bibnamefont
  {Whittle}}, \bibinfo {author} {\bibfnamefont {E.~D.}\ \bibnamefont {Hall}},
  \bibinfo {author} {\bibfnamefont {S.}~\bibnamefont {Dwyer}}, \bibinfo
  {author} {\bibfnamefont {N.}~\bibnamefont {Mavalvala}}, \bibinfo {author}
  {\bibfnamefont {V.}~\bibnamefont {Sudhir}}, \bibinfo {author} {\bibfnamefont
  {R.}~\bibnamefont {Abbott}}, \bibinfo {author} {\bibfnamefont
  {A.}~\bibnamefont {Ananyeva}}, \bibinfo {author} {\bibfnamefont
  {C.}~\bibnamefont {Austin}}, \bibinfo {author} {\bibfnamefont
  {L.}~\bibnamefont {Barsotti}}, \bibinfo {author} {\bibfnamefont
  {J.}~\bibnamefont {Betzwieser}}, \bibinfo {author} {\bibfnamefont {C.~D.}\
  \bibnamefont {Blair}}, \bibinfo {author} {\bibfnamefont {A.~F.}\ \bibnamefont
  {Brooks}}, \bibinfo {author} {\bibfnamefont {D.~D.}\ \bibnamefont {Brown}},
  \bibinfo {author} {\bibfnamefont {A.}~\bibnamefont {Buikema}}, \bibinfo
  {author} {\bibfnamefont {C.}~\bibnamefont {Cahillane}}, \emph {et~al.},\
  }\href {https://doi.org/10.1126/science.abh2634} {\bibfield  {journal}
  {\bibinfo  {journal} {Science}\ }\textbf {\bibinfo {volume} {372}},\ \bibinfo
  {pages} {1333} (\bibinfo {year} {2021})}\BibitemShut {NoStop}%
\bibitem [{\citenamefont {Chu}\ \emph {et~al.}(2017)\citenamefont {Chu},
  \citenamefont {Kharel}, \citenamefont {Renninger}, \citenamefont {Burkhart},
  \citenamefont {Frunzio}, \citenamefont {Rakich},\ and\ \citenamefont
  {Schoelkopf}}]{chuQuantumAcousticsSuperconducting2017}%
  \BibitemOpen
  \bibfield  {author} {\bibinfo {author} {\bibfnamefont {Y.}~\bibnamefont
  {Chu}}, \bibinfo {author} {\bibfnamefont {P.}~\bibnamefont {Kharel}},
  \bibinfo {author} {\bibfnamefont {W.~H.}\ \bibnamefont {Renninger}}, \bibinfo
  {author} {\bibfnamefont {L.~D.}\ \bibnamefont {Burkhart}}, \bibinfo {author}
  {\bibfnamefont {L.}~\bibnamefont {Frunzio}}, \bibinfo {author} {\bibfnamefont
  {P.~T.}\ \bibnamefont {Rakich}},\ and\ \bibinfo {author} {\bibfnamefont
  {R.~J.}\ \bibnamefont {Schoelkopf}},\ }\href
  {https://doi.org/10.1126/science.aao1511} {\bibfield  {journal} {\bibinfo
  {journal} {Science}\ }\textbf {\bibinfo {volume} {358}},\ \bibinfo {pages}
  {199} (\bibinfo {year} {2017})}\BibitemShut {NoStop}%
\bibitem [{\citenamefont {Gustafsson}\ \emph {et~al.}(2014)\citenamefont
  {Gustafsson}, \citenamefont {Aref}, \citenamefont {Kockum}, \citenamefont
  {Ekstr{\"o}m}, \citenamefont {Johansson},\ and\ \citenamefont
  {Delsing}}]{gustafssonPropagatingPhononsCoupled2014}%
  \BibitemOpen
  \bibfield  {author} {\bibinfo {author} {\bibfnamefont {M.~V.}\ \bibnamefont
  {Gustafsson}}, \bibinfo {author} {\bibfnamefont {T.}~\bibnamefont {Aref}},
  \bibinfo {author} {\bibfnamefont {A.~F.}\ \bibnamefont {Kockum}}, \bibinfo
  {author} {\bibfnamefont {M.~K.}\ \bibnamefont {Ekstr{\"o}m}}, \bibinfo
  {author} {\bibfnamefont {G.}~\bibnamefont {Johansson}},\ and\ \bibinfo
  {author} {\bibfnamefont {P.}~\bibnamefont {Delsing}},\ }\href
  {https://doi.org/10.1126/science.1257219} {\bibfield  {journal} {\bibinfo
  {journal} {Science}\ }\textbf {\bibinfo {volume} {346}},\ \bibinfo {pages}
  {207} (\bibinfo {year} {2014})}\BibitemShut {NoStop}%
\bibitem [{\citenamefont {Lee}\ \emph {et~al.}(2012)\citenamefont {Lee},
  \citenamefont {Sussman}, \citenamefont {Sprague}, \citenamefont
  {Michelberger}, \citenamefont {Reim}, \citenamefont {Nunn}, \citenamefont
  {Langford}, \citenamefont {Bustard}, \citenamefont {Jaksch},\ and\
  \citenamefont {Walmsley}}]{leeMacroscopicNonclassicalStates2012}%
  \BibitemOpen
  \bibfield  {author} {\bibinfo {author} {\bibfnamefont {K.~C.}\ \bibnamefont
  {Lee}}, \bibinfo {author} {\bibfnamefont {B.~J.}\ \bibnamefont {Sussman}},
  \bibinfo {author} {\bibfnamefont {M.~R.}\ \bibnamefont {Sprague}}, \bibinfo
  {author} {\bibfnamefont {P.}~\bibnamefont {Michelberger}}, \bibinfo {author}
  {\bibfnamefont {K.~F.}\ \bibnamefont {Reim}}, \bibinfo {author}
  {\bibfnamefont {J.}~\bibnamefont {Nunn}}, \bibinfo {author} {\bibfnamefont
  {N.~K.}\ \bibnamefont {Langford}}, \bibinfo {author} {\bibfnamefont {P.~J.}\
  \bibnamefont {Bustard}}, \bibinfo {author} {\bibfnamefont {D.}~\bibnamefont
  {Jaksch}},\ and\ \bibinfo {author} {\bibfnamefont {I.~A.}\ \bibnamefont
  {Walmsley}},\ }\href {https://doi.org/10.1038/nphoton.2011.296} {\bibfield
  {journal} {\bibinfo  {journal} {Nature Photonics}\ }\textbf {\bibinfo
  {volume} {6}},\ \bibinfo {pages} {41} (\bibinfo {year} {2012})}\BibitemShut
  {NoStop}%
\bibitem [{\citenamefont {Duan}\ \emph {et~al.}(2001)\citenamefont {Duan},
  \citenamefont {Lukin}, \citenamefont {Cirac},\ and\ \citenamefont
  {Zoller}}]{Duan2001}%
  \BibitemOpen
  \bibfield  {author} {\bibinfo {author} {\bibfnamefont {L.-M.}\ \bibnamefont
  {Duan}}, \bibinfo {author} {\bibfnamefont {M.~D.}\ \bibnamefont {Lukin}},
  \bibinfo {author} {\bibfnamefont {J.~I.}\ \bibnamefont {Cirac}},\ and\
  \bibinfo {author} {\bibfnamefont {P.}~\bibnamefont {Zoller}},\ }\href
  {https://doi.org/10.1038/35106500} {\bibfield  {journal} {\bibinfo  {journal}
  {Nature}\ }\textbf {\bibinfo {volume} {414}},\ \bibinfo {pages} {413}
  (\bibinfo {year} {2001})}\BibitemShut {NoStop}%
\bibitem [{\citenamefont {Hong}\ \emph {et~al.}(2017)\citenamefont {Hong},
  \citenamefont {Riedinger}, \citenamefont {Marinkovi{\'c}}, \citenamefont
  {Wallucks}, \citenamefont {Hofer}, \citenamefont {Norte}, \citenamefont
  {Aspelmeyer},\ and\ \citenamefont
  {Gr{\"o}blacher}}]{hongHanburyBrownTwiss2017}%
  \BibitemOpen
  \bibfield  {author} {\bibinfo {author} {\bibfnamefont {S.}~\bibnamefont
  {Hong}}, \bibinfo {author} {\bibfnamefont {R.}~\bibnamefont {Riedinger}},
  \bibinfo {author} {\bibfnamefont {I.}~\bibnamefont {Marinkovi{\'c}}},
  \bibinfo {author} {\bibfnamefont {A.}~\bibnamefont {Wallucks}}, \bibinfo
  {author} {\bibfnamefont {S.~G.}\ \bibnamefont {Hofer}}, \bibinfo {author}
  {\bibfnamefont {R.~A.}\ \bibnamefont {Norte}}, \bibinfo {author}
  {\bibfnamefont {M.}~\bibnamefont {Aspelmeyer}},\ and\ \bibinfo {author}
  {\bibfnamefont {S.}~\bibnamefont {Gr{\"o}blacher}},\ }\href
  {https://doi.org/10.1126/science.aan7939} {\bibfield  {journal} {\bibinfo
  {journal} {Science}\ }\textbf {\bibinfo {volume} {358}},\ \bibinfo {pages}
  {203} (\bibinfo {year} {2017})}\BibitemShut {NoStop}%
\bibitem [{\citenamefont {Riedinger}\ \emph {et~al.}(2016)\citenamefont
  {Riedinger}, \citenamefont {Hong}, \citenamefont {Norte}, \citenamefont
  {Slater}, \citenamefont {Shang}, \citenamefont {Krause}, \citenamefont
  {Anant}, \citenamefont {Aspelmeyer},\ and\ \citenamefont
  {Gr{\"o}blacher}}]{Riedinger2016}%
  \BibitemOpen
  \bibfield  {author} {\bibinfo {author} {\bibfnamefont {R.}~\bibnamefont
  {Riedinger}}, \bibinfo {author} {\bibfnamefont {S.}~\bibnamefont {Hong}},
  \bibinfo {author} {\bibfnamefont {R.~A.}\ \bibnamefont {Norte}}, \bibinfo
  {author} {\bibfnamefont {J.~A.}\ \bibnamefont {Slater}}, \bibinfo {author}
  {\bibfnamefont {J.}~\bibnamefont {Shang}}, \bibinfo {author} {\bibfnamefont
  {A.~G.}\ \bibnamefont {Krause}}, \bibinfo {author} {\bibfnamefont
  {V.}~\bibnamefont {Anant}}, \bibinfo {author} {\bibfnamefont
  {M.}~\bibnamefont {Aspelmeyer}},\ and\ \bibinfo {author} {\bibfnamefont
  {S.}~\bibnamefont {Gr{\"o}blacher}},\ }\href
  {https://doi.org/10.1038/nature16536} {\bibfield  {journal} {\bibinfo
  {journal} {Nature}\ }\textbf {\bibinfo {volume} {530}},\ \bibinfo {pages}
  {313} (\bibinfo {year} {2016})}\BibitemShut {NoStop}%
\bibitem [{\citenamefont {Enzian}\ \emph {et~al.}(2021)\citenamefont {Enzian},
  \citenamefont {Freisem}, \citenamefont {Price}, \citenamefont {Svela},
  \citenamefont {Clarke}, \citenamefont {Shajilal}, \citenamefont {Janousek},
  \citenamefont {Buchler}, \citenamefont {Lam},\ and\ \citenamefont
  {Vanner}}]{enzianNonGaussianMechanicalMotion2021}%
  \BibitemOpen
  \bibfield  {author} {\bibinfo {author} {\bibfnamefont {G.}~\bibnamefont
  {Enzian}}, \bibinfo {author} {\bibfnamefont {L.}~\bibnamefont {Freisem}},
  \bibinfo {author} {\bibfnamefont {J.~J.}\ \bibnamefont {Price}}, \bibinfo
  {author} {\bibfnamefont {A.~{\O}.}\ \bibnamefont {Svela}}, \bibinfo {author}
  {\bibfnamefont {J.}~\bibnamefont {Clarke}}, \bibinfo {author} {\bibfnamefont
  {B.}~\bibnamefont {Shajilal}}, \bibinfo {author} {\bibfnamefont
  {J.}~\bibnamefont {Janousek}}, \bibinfo {author} {\bibfnamefont {B.~C.}\
  \bibnamefont {Buchler}}, \bibinfo {author} {\bibfnamefont {P.~K.}\
  \bibnamefont {Lam}},\ and\ \bibinfo {author} {\bibfnamefont {M.~R.}\
  \bibnamefont {Vanner}},\ }\href
  {https://doi.org/10.1103/PhysRevLett.127.243601} {\bibfield  {journal}
  {\bibinfo  {journal} {Physical Review Letters}\ }\textbf {\bibinfo {volume}
  {127}},\ \bibinfo {pages} {243601} (\bibinfo {year} {2021})}\BibitemShut
  {NoStop}%
\bibitem [{\citenamefont {Belenchia}\ \emph {et~al.}(2018)\citenamefont
  {Belenchia}, \citenamefont {Wald}, \citenamefont {Giacomini}, \citenamefont
  {Castro-Ruiz}, \citenamefont {Brukner},\ and\ \citenamefont
  {Aspelmeyer}}]{Belenchia2018}%
  \BibitemOpen
  \bibfield  {author} {\bibinfo {author} {\bibfnamefont {A.}~\bibnamefont
  {Belenchia}}, \bibinfo {author} {\bibfnamefont {R.~M.}\ \bibnamefont {Wald}},
  \bibinfo {author} {\bibfnamefont {F.}~\bibnamefont {Giacomini}}, \bibinfo
  {author} {\bibfnamefont {E.}~\bibnamefont {Castro-Ruiz}}, \bibinfo {author}
  {\bibfnamefont {{\v{C}}.}~\bibnamefont {Brukner}},\ and\ \bibinfo {author}
  {\bibfnamefont {M.}~\bibnamefont {Aspelmeyer}},\ }\href
  {https://doi.org/10.1103/PhysRevD.98.126009} {\bibfield  {journal} {\bibinfo
  {journal} {Phys. Rev. D}\ }\textbf {\bibinfo {volume} {98}},\ \bibinfo
  {pages} {126009} (\bibinfo {year} {2018})}\BibitemShut {NoStop}%
\bibitem [{\citenamefont {Tsaturyan}\ \emph {et~al.}(2017)\citenamefont
  {Tsaturyan}, \citenamefont {Barg}, \citenamefont {Polzik},\ and\
  \citenamefont {Schliesser}}]{Tsaturyan2017}%
  \BibitemOpen
  \bibfield  {author} {\bibinfo {author} {\bibfnamefont {Y.}~\bibnamefont
  {Tsaturyan}}, \bibinfo {author} {\bibfnamefont {A.}~\bibnamefont {Barg}},
  \bibinfo {author} {\bibfnamefont {E.~S.}\ \bibnamefont {Polzik}},\ and\
  \bibinfo {author} {\bibfnamefont {A.}~\bibnamefont {Schliesser}},\ }\href
  {https://doi.org/10.1038/nnano.2017.101} {\bibfield  {journal} {\bibinfo
  {journal} {Nature Nanotechnology}\ }\textbf {\bibinfo {volume} {12}},\
  \bibinfo {pages} {776} (\bibinfo {year} {2017})}\BibitemShut {NoStop}%
\bibitem [{Note1()}]{Note1}%
  \BibitemOpen
  \bibinfo {note} {Fabricated by FiveNine Optics}\BibitemShut {NoStop}%
\bibitem [{\citenamefont {B\o{}rkje}\ \emph {et~al.}(2011)\citenamefont
  {B\o{}rkje}, \citenamefont {Nunnenkamp},\ and\ \citenamefont
  {Girvin}}]{PhysRevLett.107.123601}%
  \BibitemOpen
  \bibfield  {author} {\bibinfo {author} {\bibfnamefont {K.}~\bibnamefont
  {B\o{}rkje}}, \bibinfo {author} {\bibfnamefont {A.}~\bibnamefont
  {Nunnenkamp}},\ and\ \bibinfo {author} {\bibfnamefont {S.~M.}\ \bibnamefont
  {Girvin}},\ }\href {https://doi.org/10.1103/PhysRevLett.107.123601}
  {\bibfield  {journal} {\bibinfo  {journal} {Phys. Rev. Lett.}\ }\textbf
  {\bibinfo {volume} {107}},\ \bibinfo {pages} {123601} (\bibinfo {year}
  {2011})}\BibitemShut {NoStop}%
\bibitem [{\citenamefont {Vanner}\ \emph {et~al.}(2013)\citenamefont {Vanner},
  \citenamefont {Aspelmeyer},\ and\ \citenamefont
  {Kim}}]{PhysRevLett.110.010504}%
  \BibitemOpen
  \bibfield  {author} {\bibinfo {author} {\bibfnamefont {M.~R.}\ \bibnamefont
  {Vanner}}, \bibinfo {author} {\bibfnamefont {M.}~\bibnamefont {Aspelmeyer}},\
  and\ \bibinfo {author} {\bibfnamefont {M.~S.}\ \bibnamefont {Kim}},\ }\href
  {https://doi.org/10.1103/PhysRevLett.110.010504} {\bibfield  {journal}
  {\bibinfo  {journal} {Phys. Rev. Lett.}\ }\textbf {\bibinfo {volume} {110}},\
  \bibinfo {pages} {010504} (\bibinfo {year} {2013})}\BibitemShut {NoStop}%
\bibitem [{\citenamefont {Galland}\ \emph {et~al.}(2014)\citenamefont
  {Galland}, \citenamefont {Sangouard}, \citenamefont {Piro}, \citenamefont
  {Gisin},\ and\ \citenamefont {Kippenberg}}]{galland}%
  \BibitemOpen
  \bibfield  {author} {\bibinfo {author} {\bibfnamefont {C.}~\bibnamefont
  {Galland}}, \bibinfo {author} {\bibfnamefont {N.}~\bibnamefont {Sangouard}},
  \bibinfo {author} {\bibfnamefont {N.}~\bibnamefont {Piro}}, \bibinfo {author}
  {\bibfnamefont {N.}~\bibnamefont {Gisin}},\ and\ \bibinfo {author}
  {\bibfnamefont {T.~J.}\ \bibnamefont {Kippenberg}},\ }\href
  {https://doi.org/10.1103/PhysRevLett.112.143602} {\bibfield  {journal}
  {\bibinfo  {journal} {Physical Review Letters}\ }\textbf {\bibinfo {volume}
  {112}},\ \bibinfo {pages} {143602} (\bibinfo {year} {2014})}\BibitemShut
  {NoStop}%
\bibitem [{\citenamefont {Dideriksen}\ \emph {et~al.}(2021)\citenamefont
  {Dideriksen}, \citenamefont {Schmieg}, \citenamefont {Zugenmaier},\ and\
  \citenamefont {Polzik}}]{dideriksenRoomtemperatureSinglephotonSource2021a}%
  \BibitemOpen
  \bibfield  {author} {\bibinfo {author} {\bibfnamefont {K.~B.}\ \bibnamefont
  {Dideriksen}}, \bibinfo {author} {\bibfnamefont {R.}~\bibnamefont {Schmieg}},
  \bibinfo {author} {\bibfnamefont {M.}~\bibnamefont {Zugenmaier}},\ and\
  \bibinfo {author} {\bibfnamefont {E.~S.}\ \bibnamefont {Polzik}},\ }\href
  {https://doi.org/10.1038/s41467-021-24033-8} {\bibfield  {journal} {\bibinfo
  {journal} {Nature Communications}\ }\textbf {\bibinfo {volume} {12}},\
  \bibinfo {pages} {3699} (\bibinfo {year} {2021})}\BibitemShut {NoStop}%
\bibitem [{\citenamefont {Corzo}\ \emph {et~al.}(2019)\citenamefont {Corzo},
  \citenamefont {Raskop}, \citenamefont {Chandra}, \citenamefont {Sheremet},
  \citenamefont {Gouraud},\ and\ \citenamefont
  {Laurat}}]{corzoWaveguidecoupledSingleCollective2019}%
  \BibitemOpen
  \bibfield  {author} {\bibinfo {author} {\bibfnamefont {N.~V.}\ \bibnamefont
  {Corzo}}, \bibinfo {author} {\bibfnamefont {J.}~\bibnamefont {Raskop}},
  \bibinfo {author} {\bibfnamefont {A.}~\bibnamefont {Chandra}}, \bibinfo
  {author} {\bibfnamefont {A.~S.}\ \bibnamefont {Sheremet}}, \bibinfo {author}
  {\bibfnamefont {B.}~\bibnamefont {Gouraud}},\ and\ \bibinfo {author}
  {\bibfnamefont {J.}~\bibnamefont {Laurat}},\ }\href
  {https://doi.org/10.1038/s41586-019-0902-3} {\bibfield  {journal} {\bibinfo
  {journal} {Nature}\ }\textbf {\bibinfo {volume} {566}},\ \bibinfo {pages}
  {359} (\bibinfo {year} {2019})}\BibitemShut {NoStop}%
\bibitem [{\citenamefont {Aspelmeyer}\ \emph {et~al.}(2014)\citenamefont
  {Aspelmeyer}, \citenamefont {Kippenberg},\ and\ \citenamefont
  {Marquardt}}]{Aspelmeyer2014}%
  \BibitemOpen
  \bibfield  {author} {\bibinfo {author} {\bibfnamefont {M.}~\bibnamefont
  {Aspelmeyer}}, \bibinfo {author} {\bibfnamefont {T.~J.}\ \bibnamefont
  {Kippenberg}},\ and\ \bibinfo {author} {\bibfnamefont {F.}~\bibnamefont
  {Marquardt}},\ }\href {https://doi.org/10.1103/RevModPhys.86.1391} {\bibfield
   {journal} {\bibinfo  {journal} {Reviews of Modern Physics}\ }\textbf
  {\bibinfo {volume} {86}},\ \bibinfo {pages} {1391} (\bibinfo {year}
  {2014})}\BibitemShut {NoStop}%
\bibitem [{\citenamefont {Reid}\ and\ \citenamefont
  {Walls}(1986)}]{reidViolationsClassicalInequalities1986}%
  \BibitemOpen
  \bibfield  {author} {\bibinfo {author} {\bibfnamefont {M.~D.}\ \bibnamefont
  {Reid}}\ and\ \bibinfo {author} {\bibfnamefont {D.~F.}\ \bibnamefont
  {Walls}},\ }\href {https://doi.org/10.1103/PhysRevA.34.1260} {\bibfield
  {journal} {\bibinfo  {journal} {Physical Review A}\ }\textbf {\bibinfo
  {volume} {34}},\ \bibinfo {pages} {1260} (\bibinfo {year}
  {1986})}\BibitemShut {NoStop}%
\bibitem [{sup()}]{suppmat}%
  \BibitemOpen
  \href@noop {} {}\bibinfo {note} {See Supplemental Material at [URL will be
  inserted by publisher] for technical implementation details.}\BibitemShut
  {Stop}%
\bibitem [{\citenamefont {Galinskiy}\ \emph
  {et~al.}(2020{\natexlab{a}})\citenamefont {Galinskiy}, \citenamefont
  {Tsaturyan}, \citenamefont {Parniak},\ and\ \citenamefont
  {Polzik}}]{ourOptica}%
  \BibitemOpen
  \bibfield  {author} {\bibinfo {author} {\bibfnamefont {I.}~\bibnamefont
  {Galinskiy}}, \bibinfo {author} {\bibfnamefont {Y.}~\bibnamefont
  {Tsaturyan}}, \bibinfo {author} {\bibfnamefont {M.}~\bibnamefont {Parniak}},\
  and\ \bibinfo {author} {\bibfnamefont {E.~S.}\ \bibnamefont {Polzik}},\
  }\href {https://doi.org/10.1364/OPTICA.390939} {\bibfield  {journal}
  {\bibinfo  {journal} {Optica}\ }\textbf {\bibinfo {volume} {7}},\ \bibinfo
  {pages} {718} (\bibinfo {year} {2020}{\natexlab{a}})}\BibitemShut {NoStop}%
\bibitem [{\citenamefont {Parniak}\ \emph {et~al.}(2021)\citenamefont
  {Parniak}, \citenamefont {Galinskiy}, \citenamefont {Zwettler},\ and\
  \citenamefont {Polzik}}]{delaylinepaper}%
  \BibitemOpen
  \bibfield  {author} {\bibinfo {author} {\bibfnamefont {M.}~\bibnamefont
  {Parniak}}, \bibinfo {author} {\bibfnamefont {I.}~\bibnamefont {Galinskiy}},
  \bibinfo {author} {\bibfnamefont {T.}~\bibnamefont {Zwettler}},\ and\
  \bibinfo {author} {\bibfnamefont {E.~S.}\ \bibnamefont {Polzik}},\ }\href
  {https://doi.org/10.1364/OE.415942} {\bibfield  {journal} {\bibinfo
  {journal} {Optics Express}\ }\textbf {\bibinfo {volume} {29}},\ \bibinfo
  {pages} {6935} (\bibinfo {year} {2021})}\BibitemShut {NoStop}%
\bibitem [{\citenamefont {Mathiassen}(2019)}]{jonasMScThesis}%
  \BibitemOpen
  \bibfield  {author} {\bibinfo {author} {\bibfnamefont {J.}~\bibnamefont
  {Mathiassen}},\ }\emph {\bibinfo {title} {Characterising and {{Modelling
  Thermal Substrate Noise}} for a {{ Membrane}} in the {{Middle Optomechanical
  Cavity}}}},\ \href@noop {} {\bibinfo {type} {{{MSc}}}},\ \bibinfo  {school}
  {Niels Bohr Institute, Faculty of Science, University of Copenhagen}
  (\bibinfo {year} {03/Dec/2019})\BibitemShut {NoStop}%
\bibitem [{\citenamefont {Galinskiy}\ \emph
  {et~al.}(2020{\natexlab{b}})\citenamefont {Galinskiy}, \citenamefont
  {Tsaturyan}, \citenamefont {Parniak},\ and\ \citenamefont
  {Polzik}}]{galinskiyPhononCountingThermometry2020}%
  \BibitemOpen
  \bibfield  {author} {\bibinfo {author} {\bibfnamefont {I.}~\bibnamefont
  {Galinskiy}}, \bibinfo {author} {\bibfnamefont {Y.}~\bibnamefont
  {Tsaturyan}}, \bibinfo {author} {\bibfnamefont {M.}~\bibnamefont {Parniak}},\
  and\ \bibinfo {author} {\bibfnamefont {E.~S.}\ \bibnamefont {Polzik}},\
  }\href {https://doi.org/10.1364/OPTICA.390939} {\bibfield  {journal}
  {\bibinfo  {journal} {Optica}\ }\textbf {\bibinfo {volume} {7}},\ \bibinfo
  {pages} {718} (\bibinfo {year} {2020}{\natexlab{b}})}\BibitemShut {NoStop}%
\bibitem [{\citenamefont {Rabl}\ \emph {et~al.}(2009)\citenamefont {Rabl},
  \citenamefont {Genes}, \citenamefont {Hammerer},\ and\ \citenamefont
  {Aspelmeyer}}]{rablPhasenoiseInducedLimitations2009}%
  \BibitemOpen
  \bibfield  {author} {\bibinfo {author} {\bibfnamefont {P.}~\bibnamefont
  {Rabl}}, \bibinfo {author} {\bibfnamefont {C.}~\bibnamefont {Genes}},
  \bibinfo {author} {\bibfnamefont {K.}~\bibnamefont {Hammerer}},\ and\
  \bibinfo {author} {\bibfnamefont {M.}~\bibnamefont {Aspelmeyer}},\ }\href
  {https://doi.org/10.1103/PhysRevA.80.063819} {\bibfield  {journal} {\bibinfo
  {journal} {Physical Review A}\ }\textbf {\bibinfo {volume} {80}},\ \bibinfo
  {pages} {063819} (\bibinfo {year} {2009})}\BibitemShut {NoStop}%
\end{thebibliography}%
\bibliographystyle{apsrev4-2}


\newpage

\setcounter{figure}{0}
\setcounter{equation}{0}
\onecolumngrid
\section*{Supplementary Information}
\twocolumngrid

\renewcommand{\theequation}{S\arabic{equation}}
\renewcommand{\thefigure}{S\arabic{figure}}
\renewcommand{\thetable}{S\Roman{table}}
\renewcommand{\thesubsection}{S.\Roman{subsection}}

\section{Filter cavity system}
Our regime of operation imposes strict limits on the bandwidth of filtering required for single-photon counting. As mentioned in the main text, we tackled this challenge by constructing a system of four \fperot{} cavities, where the light scattered by the optomechanical interaction passes through each filter sequentially. 

\begin{figure}[htbp]
  \centering
  \renewcommand\sffamily{}
  \newcommand\hzz{\Hz}
\begingroup%
  \makeatletter%
  \providecommand\color[2][]{%
    \errmessage{(Inkscape) Color is used for the text in Inkscape, but the package 'color.sty' is not loaded}%
    \renewcommand\color[2][]{}%
  }%
  \providecommand\transparent[1]{%
    \errmessage{(Inkscape) Transparency is used (non-zero) for the text in Inkscape, but the package 'transparent.sty' is not loaded}%
    \renewcommand\transparent[1]{}%
  }%
  \providecommand\rotatebox[2]{#2}%
  \newcommand*\fsize{\dimexpr\f@size pt\relax}%
  \newcommand*\lineheight[1]{\fontsize{\fsize}{#1\fsize}\selectfont}%
  \ifx\svgwidth\undefined%
    \setlength{\unitlength}{243.77952756bp}%
    \ifx\svgscale\undefined%
      \relax%
    \else%
      \setlength{\unitlength}{\unitlength * \real{\svgscale}}%
    \fi%
  \else%
    \setlength{\unitlength}{\svgwidth}%
  \fi%
  \global\let\svgwidth\undefined%
  \global\let\svgscale\undefined%
  \makeatother%
  \begin{picture}(1,0.84883721)%
    \lineheight{1}%
    \setlength\tabcolsep{0pt}%
    \put(0,0){\includegraphics[width=\unitlength,page=1]{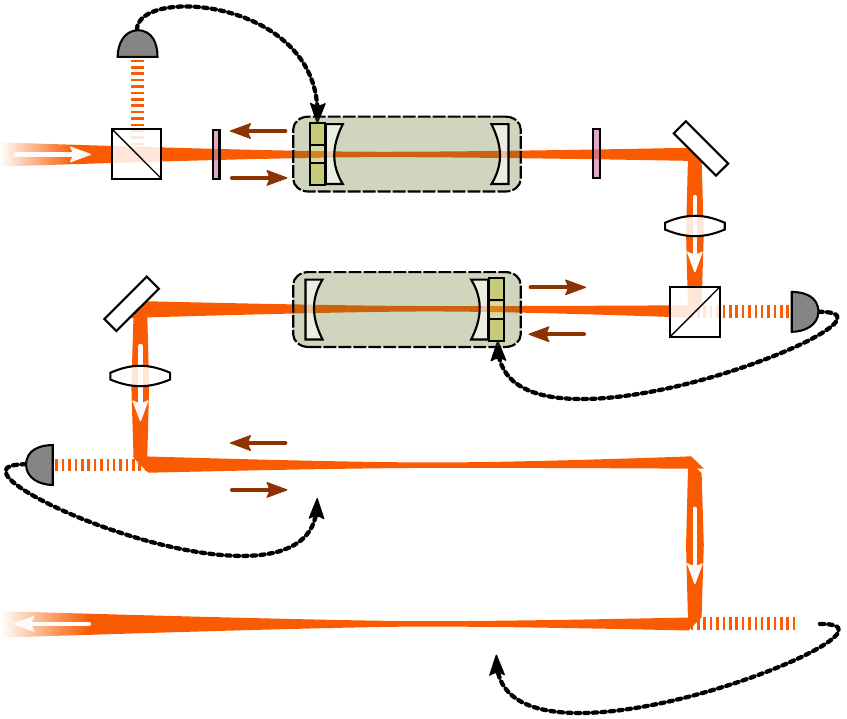}}%
    \put(0.48026125,0.57826747){\color[rgb]{0.00784314,0.00784314,0.00784314}\makebox(0,0)[t]{\lineheight{0}\smash{\begin{tabular}[t]{c}$\frac{\lambda}{4}$\end{tabular}}}}%
    \put(0.00668269,0.61549619){\color[rgb]{0.00784314,0.00784314,0.00784314}\makebox(0,0)[lt]{\lineheight{0}\smash{\begin{tabular}[t]{l}input\end{tabular}}}}%
    \put(0.00736295,0.14677895){\color[rgb]{0.00784314,0.00784314,0.00784314}\makebox(0,0)[lt]{\lineheight{0}\smash{\begin{tabular}[t]{l}output\end{tabular}}}}%
    \put(0,0){\includegraphics[width=\unitlength,page=2]{filter_system_annotated.pdf}}%
  \end{picture}%
\endgroup%

    \caption{Schematic of the 4-cavity filter system. Light that has interacted with the optomechanical system enters the filters on the top left, and propagates sequentially through all the cavities. For locking purposes, reflection from each cavity is monitored by photodetectors and is electronically stabilized to be as close to zero as possible (i.e. as close to resonance as possible)}
  \label{fig:filters}
\end{figure}

As we have described previously \cite{ourOptica}, each filter consists of two equally reflective mirrors, leading to an overall finesse of approximately \num{8800}, separated by a \qty{60}{\cm} invar spacer. These cavities are placed in a vacuum system to reduce atmospheric pressure effects and optical absorption in air. The in-series arrangement, shown schematically in \fref{fig:filters}, compounds the individual \qty{30}{\kHz} Lorentzian transfer functions into an asymptotically 8-order effective response, providing ample suppression of nearly all undesired photons. 

The choice of such a narrow linewidth was dictated by the need to suppress photons scattered by out-of-bandgap mechanical modes in close proximity to the fundamental mechanical mode. As a welcome side-effect, the rejection of unscattered pump light is extremely high and is expected to be greater than \qty{155}{\dB} \cite{ourOptica}. 

In the current implementation, the filter system displays an end-to-end signal transmission of approximately \perc{30}, dominated by losses due to the roughness of cavity mirror substrates and coupling imperfections.

\section{Phase noise of light}
\begin{figure}[htbp]
  \centering
  \renewcommand\sffamily{}
  \newcommand\hzz{\Hz}
\begingroup%
  \makeatletter%
  \providecommand\color[2][]{%
    \errmessage{(Inkscape) Color is used for the text in Inkscape, but the package 'color.sty' is not loaded}%
    \renewcommand\color[2][]{}%
  }%
  \providecommand\transparent[1]{%
    \errmessage{(Inkscape) Transparency is used (non-zero) for the text in Inkscape, but the package 'transparent.sty' is not loaded}%
    \renewcommand\transparent[1]{}%
  }%
  \providecommand\rotatebox[2]{#2}%
  \newcommand*\fsize{\dimexpr\f@size pt\relax}%
  \newcommand*\lineheight[1]{\fontsize{\fsize}{#1\fsize}\selectfont}%
  \ifx\svgwidth\undefined%
    \setlength{\unitlength}{243.77952756bp}%
    \ifx\svgscale\undefined%
      \relax%
    \else%
      \setlength{\unitlength}{\unitlength * \real{\svgscale}}%
    \fi%
  \else%
    \setlength{\unitlength}{\svgwidth}%
  \fi%
  \global\let\svgwidth\undefined%
  \global\let\svgscale\undefined%
  \makeatother%
  \begin{picture}(1,0.81395349)%
    \lineheight{1}%
    \setlength\tabcolsep{0pt}%
    \put(0,0){\includegraphics[width=\unitlength,page=1]{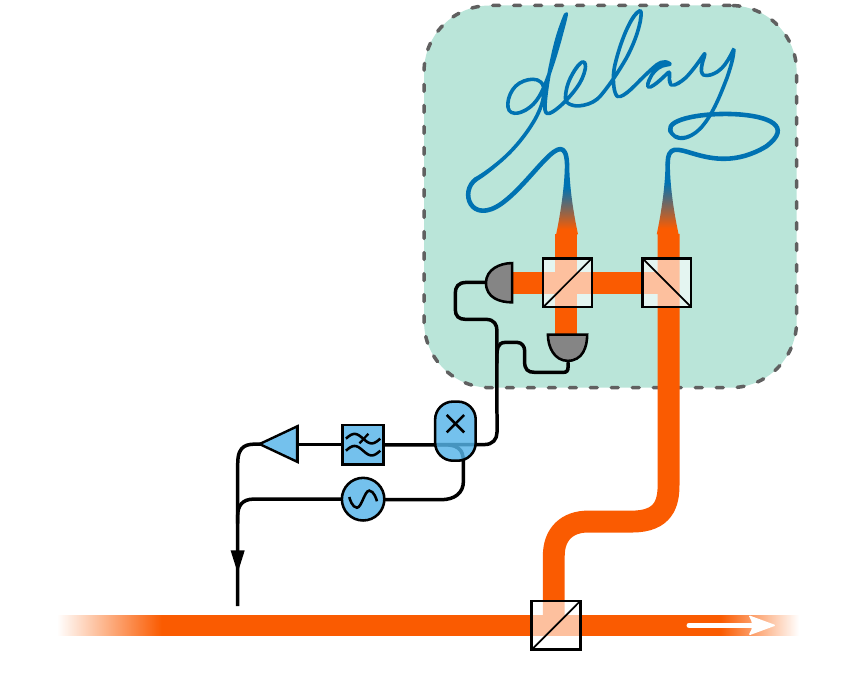}}%
    \put(0.30034846,0.31838282){\color[rgb]{0,0,0}\makebox(0,0)[lt]{\lineheight{1.25}\smash{\begin{tabular}[t]{l}PID\end{tabular}}}}%
    \put(0,0){\includegraphics[width=\unitlength,page=2]{single_loop_with_laser.pdf}}%
    \put(0.23848077,0.002976){\color[rgb]{0,0,0}\makebox(0,0)[lt]{\lineheight{1.25}\smash{\begin{tabular}[t]{l}EOM\end{tabular}}}}%
    \put(0,0){\includegraphics[width=\unitlength,page=3]{single_loop_with_laser.pdf}}%
  \end{picture}%
\endgroup%

    \caption{Schematic of phase noise cancellation using a delay line. Laser light, coming from the left, is sampled by a beamsplitter into a highly unbalanced Mach-Zehnder interferometer which transduces phase noise at \qty{1.4}{\MHz} into detectable intensity noise. This signal is bandpassed and phase-shifted using a mix-lowpass-mix electronic chain and is applied to the EOM in order to close the feedback loop.}
  \label{fig:delay-line-loop}
\end{figure}

To address phase noise coming from the laser system, we implemented a real-time feedback system based on an unbalanced Mach-Zehnder interferometer, described in detail in \cite{delaylinepaper}. A highly unbalanced interferometer, where the arms have an optical path difference of $\approx \qty{70}{\m}$ (\qty{50}{\m} of optical fiber), is used to transduce phase noise of light into a measurable intensity signal at a frequency of \qty{1.4}{\MHz}. As shown in \fref{fig:delay-line-loop}, the measured signal is then used to apply feedback on an electro-optic modulator, efficiently cancelling the phase fluctuations at that frequency. 

The overall result of our scheme is a reduction of laser phase noise by more than \qty{8}{\dB} with respect to our laser's intrinsic noise, down to a phase noise spectral density of \qty{-164}{\dBc\per\Hz} around our frequency of interest \cite{delaylinepaper}.

\section{Thermal noise of mirrors}
Mirror noise, i.e. thermal motion of reflective surfaces of the optomechanical cavity's mirrors, has an effect similar to that of external phase noise \cite{rablPhasenoiseInducedLimitations2009}. Even at liquid-helium (\qty{4}{\K}) temperatures of our cryostat and a cavity assembly temperature of \qtyrange{6}{9}{\K}, this motion is substantial enough to introduce practical limits on optical cooling and reading. 

In our design, we work around the mirror noise limitation by engineering the key dimensions of our mirrors to avoid any frequency-domain overlap between the mechanical modes of the mirrors and of the membrane's main mode. Specifically, FEM simulation \cite{jonasMScThesis} showed that a cylindrical fused-silica mirror of \qty{5}{\mm} diameter and approximately \qty{1}{\mm} thickness yields a mode structure with no substantial overlap between mirror modes and the optomechanical mode of interest. This optimized geometry yields a limit of optomechanical cooling, i.e. minimum achievable phonon number, of $\approx \num{0.2}$ for our system.

\section{Construction of optomechanical cavity}

Our optomechanical cavity is constructed out of oxygen-free copper to maximize thermal conductivity at cryogenic temperatures. The cavity is mounted on a liquid-helium flow cryostat, which provides the necessary cooling at a relatively low level of vibration. However, given the mechanical sensitivity of our system, we also suspended our cavity on four stainless steel springs, which create an effective \qty{10}{\Hz} mass-on-spring oscillator, rejecting thus external vibrations due to e.g. helium flow fluctuations or environmental noise. 

\begin{figure}[htbp]
  \centering
  \renewcommand\sffamily{}
  \newcommand\hzz{\Hz}
\begingroup%
  \makeatletter%
  \providecommand\color[2][]{%
    \errmessage{(Inkscape) Color is used for the text in Inkscape, but the package 'color.sty' is not loaded}%
    \renewcommand\color[2][]{}%
  }%
  \providecommand\transparent[1]{%
    \errmessage{(Inkscape) Transparency is used (non-zero) for the text in Inkscape, but the package 'transparent.sty' is not loaded}%
    \renewcommand\transparent[1]{}%
  }%
  \providecommand\rotatebox[2]{#2}%
  \newcommand*\fsize{\dimexpr\f@size pt\relax}%
  \newcommand*\lineheight[1]{\fontsize{\fsize}{#1\fsize}\selectfont}%
  \ifx\svgwidth\undefined%
    \setlength{\unitlength}{243.77952756bp}%
    \ifx\svgscale\undefined%
      \relax%
    \else%
      \setlength{\unitlength}{\unitlength * \real{\svgscale}}%
    \fi%
  \else%
    \setlength{\unitlength}{\svgwidth}%
  \fi%
  \global\let\svgwidth\undefined%
  \global\let\svgscale\undefined%
  \makeatother%
  \begin{picture}(1,0.75581395)%
    \lineheight{1}%
    \setlength\tabcolsep{0pt}%
    \put(0,0){\includegraphics[width=\unitlength,page=1]{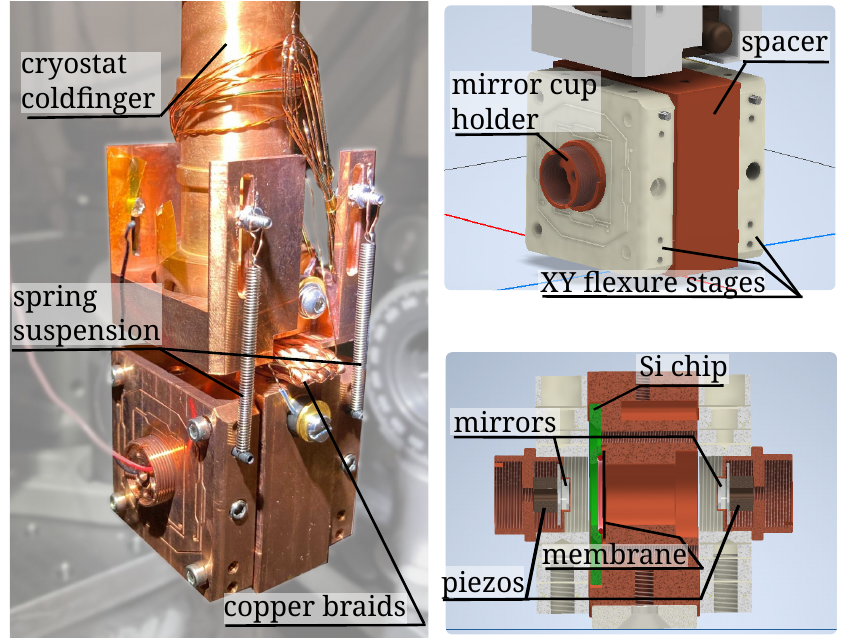}}%
    \put(0.02953231,0.70575308){\color[rgb]{0,0,0}\makebox(0,0)[lt]{\lineheight{0}\smash{\begin{tabular}[t]{l}$\textbf{(a)}$\end{tabular}}}}%
    \put(0.56065546,0.70575308){\color[rgb]{0,0,0}\makebox(0,0)[t]{\lineheight{0}\smash{\begin{tabular}[t]{c}$\textbf{(b)}$\end{tabular}}}}%
    \put(0.56065546,0.29383083){\color[rgb]{0,0,0}\makebox(0,0)[t]{\lineheight{0}\smash{\begin{tabular}[t]{c}$\textbf{(c)}$\end{tabular}}}}%
  \end{picture}%
\endgroup%

  \caption{Optomechanical cavity suspended on springs inside the helium flow cryostat (a), with a schematic representation (b), and a cross-section (c), showing the positioning of the membrane chip, mirrors, and piezo actuators. The radiation shield and vacuum chamber enclosure were removed for illustration purposes. In panel (c), the two thin mirrors (white) are located on the sides of the central copper block, while the membrane (black) is held in position by a copper clamp (green). Piezo actuators (dark grey) act on the mirrors to adjust either the optical resonance frequency (when moving in opposite directions) or the relative position between the optical standing wave and the membrane (when moving in the same direction).}
  \label{fig:om-cavity-combined}
\end{figure}

To maintain thermal conduction to the cryostat, we then added a flexible copper braid between the cold finger and the optomechanical assembly, as shown in \fref{fig:om-cavity-combined}. Due to space constraints, we opted to use a braid with a cross-section of \qty{6}{\mm^2} and an approximate length of \qty{60}{\mm}. While this limits the thermalization rate of our assembly at room temperature, the increased thermal conductivity of copper at cryogenic temperatures allows the assembly to thermalize at approximately \qtyrange{6}{9}{\K}. 

Finally, we incorporate two piezo actuators acting on both mirrors of the cavity. When the mirrors are displaced in the same direction, the position of the optical standing wave is shifted with respect to the membrane. On the other hand, when the mirrors are moved in opposite directions, the length of the cavity is changed, tuning therefore its resonance frequency for the purposes of locking.

\begin{table*}[tb!]
    \centering
    \begin{tabularx}{\textwidth}{lX}
    \hline\hline
    \: Symbol \: & Meaning \\
    \hline
        $\kappa$ & Optical decay rate (optical linewidth) \\
        $\Om$ & Mechanical frequency\\
        $\Delta$ & Detuning of coherent drive wrt. optical resonance\\
        $\gnot$ & Single-photon optomechanical coupling rate\\
         $\ncav^\text{w/r}$ & Number of coherent drive photons in the optical resonator during write/read operations, respectively\\
         $A_+$, $A_-$ & Stokes and anti-Stokes scattering rates, respectively\\
         $\Gm$ & Mechanical decay rate\\
         \hline\hline
    \end{tabularx}
    \caption{Definitions of symbols used in this manuscript.}
    \label{tab:definitions}
\end{table*}

\section{Photon rates}

\begin{figure}[tb]
    \centering
    \renewcommand\sffamily{}
    \includegraphics[width=1\columnwidth]{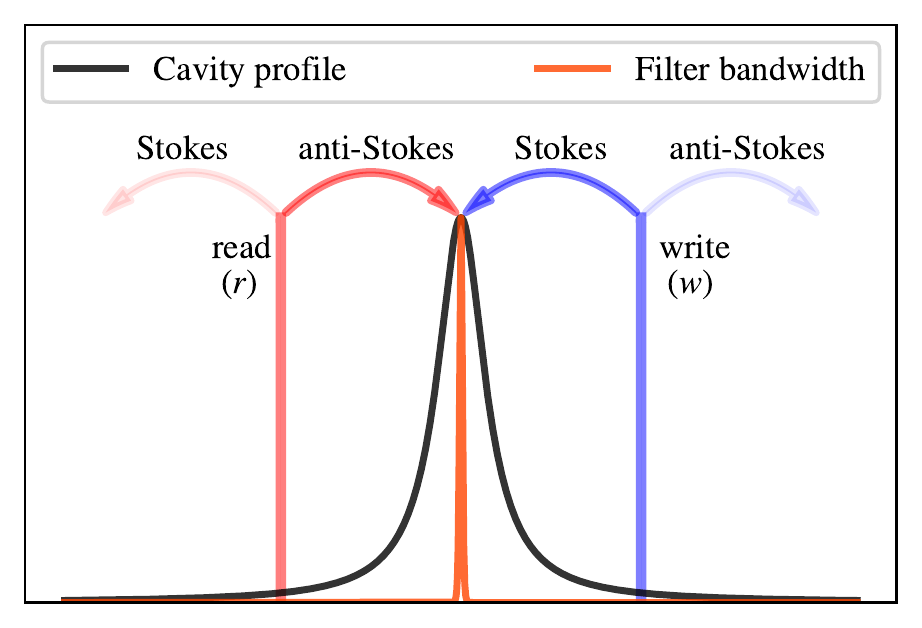}
    \caption{Sidebands in the experiment. Each drive field is scattered by mechanical motion into two sidebands at distances of $\pm \Omega$. Scattered light resonant with the optomechanical cavity is strongly preferred, but we still take into account the other sidebands in our calculation. However, only photons falling into the very narrow filter bandwidth are at the end detected by our single-photon counter.}
  \label{fig:sideb}
\end{figure}
Here we detail the calculation that leads to a theory prediction and matching for photon rates and associated phonon occupation.

Whenever the optomechanical system is illuminated by a coherent drive, photons of the drive will be scattered up in frequency, and the oscillator will move down on the Fock-state ladder (Anti-Stokes, $A_-$) or the photon will be scattered down and the oscillator will receive one phonon (Stokes, $A_+$), with the following scattering rate:
\begin{equation}
 A_\pm  = g_0^2 \ncav \frac{\kappa}{(\Delta\mp\Om)^2+\kappa^2/4} \:,
\end{equation}
with symbol definitions given in Table \ref{tab:definitions}. In our experiment, we alternate between two drive fields: reading (cooling) with detuning $\Delta_r=-\Om$, and writing (excitation) with detuning $\Delta_w=+\Om$. Given this, we can calculate the transition rates $A_\pm$ at all times, in the presence of both write and read light treated as independent drives.

With these transition rates established, we have the following equation for the evolution of the phonon number in the membrane
\begin{equation}
\frac{\mathrm{d}\bar{n}}{\mathrm{d}t} = (\bar{n}+1)(A_++\nt \Gm) - \bar{n}(A_-+(\nt+1) \Gm)
\end{equation}
In our simulation, we solve this simple equation numerically, taking into account independently measured laser light powers $P_\text{w/r}$ leaking out of the cavity. In the experiment, we measure the light in the cooling-only sequence, and then in the full sequence, to establish the power levels of each driver. We compensate for the photodiode electronic offset and for the offset due to the lock light. Through that, $\ncav^\text{w/r} \propto P_\text{w/r}$ can be determined, as the outgoing light is a direct measure of the intra-cavity photon number. The proportionality coefficient between the intra-cavity photon number and the incident laser power is left as a free parameter in matching the simulation to experimental data. Note that effectively we thus match the optomechanical coupling $g=g_0\sqrt{\ncav}$.
For the initial condition, we take the steady-state occupation of the oscillator being driven by the steady-state read laser light power during "idle" intervals:
\begin{equation}
  \nbar(0) = \frac{A_+ + \bar{n}_\mathrm{th} \Gm}{A_- - A_+ + \Gm}
\end{equation}
where $A_\pm$ are taken at $\ncav^r \propto P^r_\mathrm{steady}$ and $\ncav^w=0$. 

Thanks to our narrowband filtering system that consists of four \qty{30}{\kHz} \fperot cavities in series \cite{ourOptica}, we only detect photons that are allowed by the filter linewidth: Stokes photons from the write laser and anti-Stokes photons from the read laser, while completely disregarding photons from the other sidebands, i.e. Stokes photons from read light and anti-Stokes photons from the write laser light. This means that photon rates coming out of the optomechanical system will be given by

\begin{alignat}{1}
    \GS(t) = &\gnot^2 \cdot \ncav^w(t) \cdot \frac{\kappa}{(\Delta_w - \Om)^2+\kappa^2/4} \cdot (\nbar(t)+1) \\
    \GAS(t) = &\gnot^2 \cdot \ncav^r(t) \cdot \frac{\kappa}{(\Delta_r + \Om)^2+\kappa^2/4} \cdot \nbar(t)
\end{alignat}

Finally, to obtain expected detection rates as seen by the single-photon counter, we need to consider the action of the narrowband filtering system on the temporal dynamics of the photon trace. To do this, we simply convolve the rates $\GS(t)$, $\GAS(t)$ time-domain response function of the filtering system, i.e.:
\begin{equation}
  \Gamma(t) \rightarrow \int_{-\infty}^\infty \Gamma(t') \xi(t-t')\mathrm{d}t' \:,
\end{equation}
with
\begin{equation}
  \xi(t) = \frac{1}{6} \exp(-\kappa_f t) \kappa_f^4 \cdot t^3 \cdot \theta(t) \:,
\end{equation}
where $\kappa_f$ is the filter bandwidth of approx. $2\pi\times\qty{30}{kHz}$ and $\theta(t)$ is the Heaviside step function. The particular form of  the transfer function is due to the presence of the four filter cavities in series.

\section{Data analysis}

\begin{figure*}[tb]
    \centering
    \renewcommand\sffamily{}
    \includegraphics[width=0.8\textwidth]{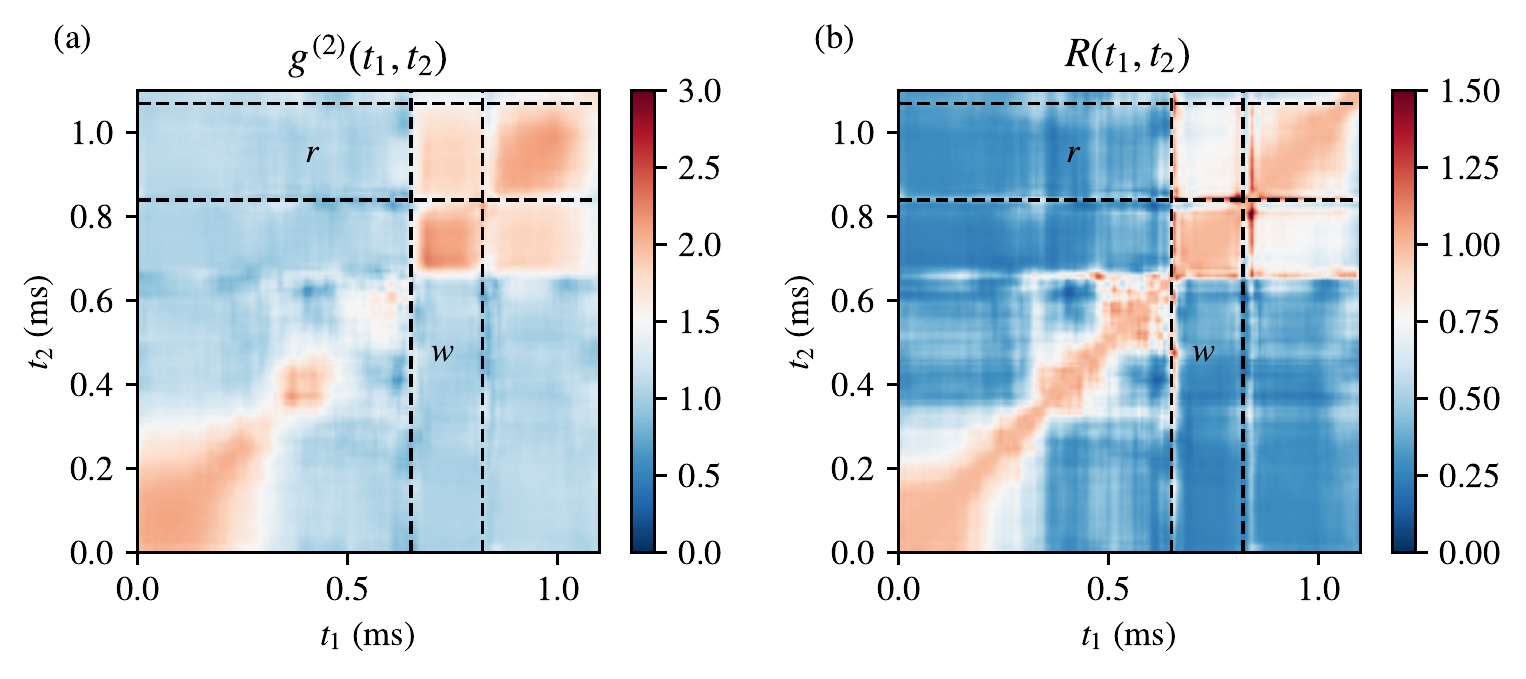}
    \caption{Experimental data presenting values of the correlation function $\gtwo$ and the \cauchy parameter $R$ for an experiment with high-probability write, generating about 10 phonons in the mechanical mode per each shot. Panel (a) presents the $\gtwo$, panel (b) - the $R$. The off-diagonal lobes in (a) and (b) outlined by the dashed lines show the correlated regions of the distributions. Distance between dashed lines represented the selected optimal mode duration. Both $r$ and $w$ modes are shifted around to obtain the presented maps.}
  \label{fig:exp-cauchy-hp}
\end{figure*}

Data analysis starts with a series of experimental shots, each containing photon timestamps. To calculate correlations, we select two time 
windows of duration $\tau$, delimited by times $(t_\text{1,2}-\tau/2, t_\text{1,2}+\tau/2)$ in each shot. The correlations between two times $t_1$ and $t_2$ are then calculated as:
\begin{equation}
  \gtwo(t_1,t_2)=\frac{\langle n(t_1) n(t_2)\rangle_\tau}{\langle n(t_1)\rangle_\tau \langle n(t_2)\rangle_\tau}
\end{equation}
under the assumption that those time intervals of length $\tau$ do not overlap. If there is a finite overlap, we divide the time slots into 
overlapping (length $\tau_\ov$, mean position $t_\ov$) and non-overlapping slots (lengths $\tau_\nv^{(1/2)}$, $t_\nv^{(1/2)}$). The lengths and mean time-slot positions are calculated for each situation. For example, in the case of full overlap, we deal with autocorrelation:
\begin{equation}
  \gtwo(t,t)=\frac{\langle n(t)(n(t)-1)\rangle_\tau}{\langle n(t)\rangle_\tau^2}
\end{equation}
In the extreme case of partial overlap, we independently calculate the overlapping and non-overlapping averages to obtain the final $\gtwo$. Care needs to be taken to correlate all combinations and consider which of the slots comes first.
In this framework, the \cauchy parameter is:
\begin{equation}
R(t_1,t_2) = \frac{[\gtwo(t_1,t_2)]^2}{\gtwo(t_1,t_1)\gtwo(t_2,t_2)}
\end{equation}
The results of this data analysis are presented in Fig.~4 of the main text for the case of low excitation strengths, leading to nonclassical dynamics. To highlight our method, we also present the same analysis for an experiment where we significantly increased the excitation probability $g_w T_w \approx 1$, shown in  \fref{fig:exp-cauchy-hp}. Here, we observe strong and distinct correlations, especially in the $\gtwo$ map, while the \cauchy{} parameter does not demonstrate non-classical features (apart from single points which occur due to low signal at some portions of the time trace, and are disregarded due to high associated uncertainty). Furthermore, in the $\gtwo$ map we observe the thermal autocorrelation of photons scattered during each process independently.

\end{document}